\documentclass[reprint,onecolumn,amsmath,amssymb,aps,nofootinbib,pagebackref=false, colorlinks=true]{revtex4-2}
\usepackage{subfigure}
\usepackage{graphicx}
\usepackage{dcolumn}
\usepackage{bm}
\usepackage{epsfig,amsmath}
\usepackage{amssymb}
\usepackage{multirow}
\usepackage{cases}
\usepackage{gensymb}
\usepackage{ulem}
\usepackage{comment}
\usepackage{orcidlink}
\definecolor{redish}{rgb}{0.7,0.2,0.0}  
\definecolor{bluish}{rgb}{0.2,0.5,0.8}

\hypersetup{linkcolor=redish,          
	citecolor=blue,        
	filecolor=magenta,      
	urlcolor=bluish}          

\DeclareFontFamily{U}{rsfs}{}         
\DeclareFontShape{U}{rsfs}{m}{n}{<5> rsfs5 <6><7> rsfs7          %
	<8><9><10><10.95><12><14.4><17.28><20.74><24.88> rsfs10}{}     %

\def \r{\rho}

\def \T{\Theta}

\def \ra{\rightarrow}

\begin{document}

\title{Transmutation Timescales for Dark Matter Induced Collapse of Compact Stars into Black Holes}

	\author{H. A. Adarsha\orcidlink{0009-0008-7781-4914
}}
\email{adarsha.mcnsmpl2023@learner.manipal.edu}
\affiliation{Manipal Centre for Natural Sciences, Manipal Academy of Higher Education, Manipal 576104, India}

  \author{Chandrachur Chakraborty\orcidlink{0000-0003-4380-3033
}}
\email{chandrachur.c@manipal.edu}
\affiliation{Manipal Centre for Natural Sciences, Manipal Academy of Higher Education, Manipal 576104, India}

\author{Sudip Bhattacharyya\orcidlink{0000-0002-6351-5808}}
\email{sudip@tifr.res.in}
\affiliation{Department of Astronomy and Astrophysics,
Tata Institute of Fundamental Research, Mumbai 400005, India}

\begin{abstract}
Ultra-heavy asymmetric dark matter (DM) particles captured by compact stars can thermalize, self-gravitate, and collapse to form an endoparasitic black hole (EBH), whose subsequent growth may transmute the host star into a black hole. The continued existence of old millisecond pulsars (MSPs) and white dwarfs (WDs) thus places powerful constraints on the DM particle mass $m_\chi$ and the DM-nucleon scattering cross-section $\sigma_{\rm n\chi}$. In this work, we derive an  analytical expression for the transmutation timescale by solving the EBH growth equation while consistently accounting for the Bondi accretion of stellar matter, Hawking evaporation, and allowing for the possibility of sustained DM feeding of the EBH in a steady-state capture regime. We also incorporate quantum mechanical effects in the baryonic accretion process by modeling particle absorption in regimes where the hydrodynamic description breaks down, thereby providing a unified treatment of EBH growth across both particle and fluid regimes. 
Adopting a physically transparent collapse criterion for both fermionic and bosonic asymmetric DM, we compute the EBH transmutation timescales for representative MSPs and WDs residing in environments with different DM densities. Although the underlying physical ingredients are broadly the same as those considered in previous studies, the present work derives updated constraints through a closed-form analytical treatment of EBH growth, together with the adopted prescription for the EBH formation timescale. Consequently, we obtain a lower critical EBH mass for sustained growth and revised transmutation timescales. Requiring the transmutation time to exceed $\sim 1~\mathrm{Gyr}$ for MSPs and $\sim 10~\mathrm{Gyr}$ for WDs, we derive revised constraints on $\sigma_{\rm n\chi}$ over the DM mass range $10^{6}\!-\!10^{14}~\mathrm{GeV}$. Notably, we show that EBHs with initial masses as small as $\sim 4 \times 10^{4}~\mathrm{kg}$ can undergo sustained growth, thereby extending the region of the DM parameter space that can be probed using compact stars.
\end{abstract}

\maketitle

\section{\label{sec1}Introduction}
The rotation curves of individual galaxies, clusters of galaxies, and many other astrophysical observations indicate the existence of a missing component of mass, referred to as dark matter (DM)~\cite{History_DM_2018}. 
There are several proposed DM candidates. As opposed to other possible constituents, such as primordial black holes (PBHs)~\cite{Carr_Kuhnel_2022, Montero-Camacho_2019} and/or massive compact halo objects (MACHOs)~\cite{Gelmini_2017}, a gravitomagnetic monopole~\cite{C_Chakraborty_22_gravmag}, DM could also be made of dark matter particles. 
Starting from the lowest mass ($\sim 10^{-30}$ eV) Fuzzy DM~\cite{Gelmini_2017}, DM particles could also include axions~\cite{Bartram_2021}, light DM, weakly interacting massive particles ($\sim 10-10^{4}$ GeV)~\cite{Meng_2021} and other massive DM particles~\cite{dasgupta_low_2021, Digman_2019} (see Fig. 1 of~\cite{Gelmini_2017} for details).

It is suggested that the interaction between DM particles and regular baryons is extremely small, except via gravity. Thus, it is still unclear how the DM particles could be detected. 
It is also unclear what the mass of the DM particles is or what the DM-nucleon scattering cross-section is during interactions with celestial objects, which can serve as DM detectors. 
Celestial objects can capture DM particles, which are supposed to be collected at the core of the said objects, depending on their mass ($m_{\chi}$) and DM-nucleon interaction cross-section ($\sigma_{\rm n\chi}$). 
Therefore, using the capture process, complementary constraints can be provided for comparison. Constraining $m_{\chi}$ and $\sigma_{\rm n\chi}$ is crucial. 
For instance, if a range of mass corresponds to a certain species of DM, then the allowed mass range can be probed, and that species can be characterized by constraining $m_{\chi}$ and $\sigma_{\rm n\chi}$. 
Fig. 3 of~\cite{JE_Kim_2010} (see also~\cite{Baer2015, CPdlH_Carlos_2020, de_Martino_2020}) provides an idea of $m_{\chi}-\sigma_{\rm n\chi}$ relation for various species of DM particles and how to constrain them by applying theoretical~\cite{Digman_2019, CHAN_2022, Bramante_2020} and observational~\cite{Meng_2021, Bartram_2021} results. 

The capture process may involve a single scattering~\cite{kouvaris_constraining_2011, mcdermott_constraints_2012} of a DM particle by a nucleon (for $m_\chi<10^6$ GeV) or multiple scatterings~\cite{bramante_multiscatter_2017,acevedo_supernovae_2019,Ilie_Levy_2021,Ilie_2024} for heavier ($m_\chi>10^6$ GeV) DM particles.
The idea of constraining the DM mass and interaction cross-section with baryons using celestial bodies dates back to Goldman and Nussinov~\cite{goldman_weakly_1989}. Using the existence of neutron stars (NSs), which capture Weakly Interacting Massive Particles (WIMPs) and subsequently form a tiny black hole (BH) called an endoparasitic black hole (EBH) that eventually transmutes the NS into a BH on a timescale that cannot be shorter than the observed ages of NSs, a range of WIMP masses was ruled out \cite{kouvaris_constraining_2011,mcdermott_constraints_2012,bramante_detecting_2014}.
Subsequently, this idea was extended by making the DM-baryon interaction cross-section a free variable to be determined by various astrophysical observations, such as the temperature-luminosity of compact stars~\cite{bertone_compact_2008, kouvaris_wimp_2008, Kouvaris_Tinyakov_2010}. 
In many of these studies, the DM cloud does not form a tiny black hole inside compact stars because they cannot self-gravitate or because of the self-annihilation of DM particles. However, in the case of non-annihilating or asymmetric DM, a DM core can form and self-gravitate, eventually collapsing into a black hole. 
By comparing the combined capture timescale and the timescale of accretion by the tiny black hole with the observed age of the hosts, the DM mass and DM-baryon interaction cross-section can be constrained~\cite{kouvaris_constraining_2011,mcdermott_constraints_2012,bell_realistic_2013,bramante_constraints_2013, Chakraborty_2024, Chakraborty_low_mass_nakedsingularity2024}. 
These studies considered that the timescale of EBH growth is solely determined by the spherical Bondi-type accretion of baryonic matter of the host. However, during accretion, EBH can capture DM particles~\cite{janish_type_2019, acevedo_supernovae_2019, steigerwald_revisiting_2022} and lose mass due to Hawking radiation~\cite{janish_type_2019, acevedo_supernovae_2019, steigerwald_revisiting_2022}. 
Several studies~\cite{mcdermott_constraints_2012, garani_new_2019} have considered all three factors and obtained constraints on the DM-baryon interaction with the single-scatter capture formalism for DM capture by stellar objects, as well as using multiple-scatter capture~\cite{steigerwald_revisiting_2022}. In the present work, we derive a closed-form analytical expression for the transmutation timescale that consistently incorporates baryonic accretion, Hawking evaporation, continued dark matter accretion, and quantum corrections to baryonic accretion in the multiple-scatter regime. The analytical formulation is complemented by numerical calculations, allowing us to validate the closed-form solution within its domain of applicability while also providing physical insight into the interplay between these processes and their impact on the transmutation timescale.

Millisecond Pulsars (MSPs), with rotational periods less than about $10$ ms, could be used as DM detectors. 
These NSs might have been spun up by the accretion of material from a companion star in a close binary system to form MSPs~\cite{alpar_new_1982,Radhakrishnan_1982}. 
Many MSPs are found in globular clusters, which are at least a billion years~\cite{Kiziltan_2010} old, with some exceptions (for example, MSP J1823-3021A~\cite{Fermi_LAT_colla_2011}). However, because DM particles may be continuously captured by MSPs or NSs, after a certain limit, these captured particles may collapse and form an EBH at the core of an MSP~\cite{baumgarte_max_surv_2021}. After formation, the EBH immediately starts to accrete matter from its host~\cite{dasgupta_low_2021}, and after a certain time, the entire MSP is transmuted to a BH, in principle. 
As the typical age ($T_{\rm MSP}$) of MSPs is at least one billion years (i.e., $T_{\rm MSP} \sim 1$ Gyr.)~\cite{Kiziltan_2010} and we can still observe them in the Universe, we cannot exclude that an EBH has already formed or going to form in their core, but sufficient time has not elapsed to transmute them into BHs. 
Although we expect the rotation of MSPs to alter the accretion rate onto the EBH, it has been shown~\cite{kouvaris_growth_2014, AHA_stall_ang_2025} that the change in accretion rate is not significant enough to affect the transmutation timescale or alter the end state. Therefore, we can apply simple spherical accretion to account for baryon accretion by EBH and calculate the effective transmutation timescale, along with DM accretion and Hawking radiation, even in the case of MSPs.
These assumptions should enable us to constrain the mass of DM particles and the DM-nucleon scattering cross-section through the ages of millisecond pulsars.

White dwarfs (WDs) form at the end of the life cycle of a relatively low-mass star. 
Similar to MSPs, WDs can also capture DM particles~\cite{mccullough_capture_2010,acevedo_supernovae_2019}, which eventually form EBH in their cores ~\cite{janish_type_2019,steigerwald_revisiting_2022}. 
Although the end state of such a WD can be a BH or a Kerr superspinar~\cite{Chakraborty_2024} or even stalled accretion~\cite{AHA_stall_ang_2025, AHA_2025_MAT}, depending on the spin period, an end state different from a BH is possible only in extreme cases. 
Therefore, regular WDs (with slow rotation and low magnetic fields) with an average age of $T_{\rm WD}\sim10$ Gyr can constrain $m_\chi$ and $\sigma_{\rm n\chi}$ in the same way MSPs do. 

In this paper, we constrain the DM particle mass ($m_{\chi}$) and the DM-nucleon scattering cross-section ($\sigma_{\rm n\chi}$) through the ages of MSPs and WDs. The paper is organized as follows. In Sec.~\ref{sec2}, we briefly discuss the multiple-scatter capture of DM particles following~\cite{Ilie_2024}. In Sec.~\ref{sec3}, we discuss the conditions for the formation of an EBH at the core of a host and calculate the formation timescale. The baryon capture rate, including quantum corrections to the accretion rate, is presented in Sec.~\ref{sec3.5}.
Once a black hole forms, it begins to accrete matter from its host. In Sec.~\ref{sec4}, we compute both the accretion and transmutation timescales. The formulation developed in Sec.~\ref{sec4} is then applied in Sec.~\ref{sec5} to MSPs and WDs to constrain $m_{\chi}$ and $\sigma_{\rm n\chi}$. We conclude in Sec.~\ref{sec6}.

\section{\label{sec2} Capture of DM particles by an astrophysical object}

DM particles with a particle mass exceeding $\sim10^6$ GeV require multiple scatterings to lose their kinetic energy and be gravitationally captured by celestial objects such as stars and compact stars. 
Because their momentum is large owing to their mass, DM particles may not lose sufficient kinetic energy in a single collision to be bound to their scatterer (material of celestial objects).
This necessitates multi-scatter capture rate ($F$) formalism~\cite{bramante_multiscatter_2017} whose closed-form is given by~\cite{Ilie_2024}
\begin{eqnarray}
 F=\frac{1}{3}\pi R^2\sqrt{\frac{6}{\pi}}\frac{\rho_{\rm DM}}{m_{\chi}\bar{v}} \left[(2\bar{v}^2 + 3v_{\rm esc}^2)S_1 - 2\bar{v}^2 S_2 -3S_3\right]
 \label{capture}
\end{eqnarray}
where,
\begin{eqnarray}
  S_1 \approx 1-\mathcal{O}(1/\tau^2),
 \\
 S_2=e^l\frac{2}{\tau^2}\left[\Sigma_1+\Sigma_N\right],
 \\
 S_3=v_{\rm esc}^2\left[S_2-\frac{\partial S_2}{\partial l}\right],
\end{eqnarray}
 $\rho_{\rm DM}$ is the density of the DM, $m_{\chi}$ is the mass of a DM particle, $\bar v=220$ km/s ($\equiv 10^{-3}c$) is the DM velocity dispersion, and $v_{\rm esc}=\sqrt{2GM/R}$ is the escape velocity of the host.\footnote{$G$ is the gravitational constant, $c$ is the speed of light in vacuum, $k$ is the Boltzmann constant and $\hbar$ is the Planck constant.}
The optical depth, $\tau \equiv 2R\sigma\rho/m_n$, quantifies the average number of collisions experienced by a DM particle as it traverses a host object of mass $M$, radius $R$, and density $\rho$. Here, $\sigma$ denotes the DM--target interaction cross-section, and $m_n$ is the mass of the target particle. It is convenient to express the optical depth as $\tau = \sigma/\sigma_{\rm crit}$, where $\sigma_{\rm crit}=m_n/(2R\rho)$ is the critical cross-section corresponding to an optical depth of unity, often referred to as the saturation cross-section~\cite{bramante_multiscatter_2017}. In the optically thin regime ($\tau \lesssim 1$), DM capture is dominated by single-scatter events~\cite{Ilie_2024}, and the capture rate is well approximated by the standard single-scatter expression~\cite{bramante_detecting_2014, Ilie_2024}. In contrast, for optically thick hosts ($\tau \gg 1$), multiple scatterings become important, and the capture process must be described using the multiscatter formalism. In this regime, the quantities $\Sigma_1$ and $\Sigma_N$ are given by~\cite{Ilie_2024}
\begin{eqnarray}
    \Sigma_1& \approx &\int_{1}^{\tau}\exp\left[-l(1-k/l)^{-N}\right] dN,\label{eq:Sigma1def}\\
    \Sigma_N &\approx &\int_{1}^{\tau} N\exp\left[-l(1-k/l)^{-N}\right] dN\label{eq:SigmaNDef}
\end{eqnarray}
where,
\begin{eqnarray}
l = \frac{3v_{\rm esc}^2}{2\bar{v}^2}\,\, , \hspace{1cm}
k = \frac{l}{2}.\frac{4m_{\chi}m_n}{(m_{\chi}+m_n)^2},
\end{eqnarray}
and, $N$ is the number of collisions in the case of multiple scattering. 
One may refer to Secs. 2 and 3.1 of~\cite{Ilie_2024} for a detailed discussion on the multiple and single-scattering limits.

\section{\label{sec3} Black hole formation inside an astrophysical object }
Celestial objects, such as stars and compact stars, located in regions of relatively large DM density (in the Galactic bulge, disk, and possibly globular clusters) begin capturing DM particles soon after formation. We assume that the capture process occurs continuously thereafter, i.e., after $t=0$. 
The captured DM particles undergo multiple collisions with the host matter (baryons) and thermalize at the core of the host~\cite{mcdermott_constraints_2012}.
For neutron stars, the thermalization time ($t_{\rm th}^{\rm NS}$) is the time required for DM particles to attain a kinetic energy equal to the thermal energy of baryons (neutrons), which is given by 

\begin{equation}
     t_{\rm th}^{\rm NS}\approx 10^3\; {\rm yrs} \frac{m_\chi/m_n}{(1+m_\chi/m_n)^2}\left(\frac{10^{-45}{\rm cm}^2}{\sigma_{\rm n\chi}}\right)+1.7\; {\rm yrs}\left(\frac{m_\chi}{10^{10}\; {\rm GeV}}\right)\left(\frac{10^{-45}{\rm cm}^2}{\sigma_{\rm n\chi}}\right)
    \label{thermNS}
\end{equation}
following~\cite{Bertoni_2013, Bell_2024} where the temperature of NS core is taken as $10^5$ K, and $\sigma_{\rm n\chi}$ is the DM-nucleon interaction cross-section.
For WDs, the thermalization time ($t_{\rm th}^{\rm WD}$) is given in two parts: the first ($t_{\rm th}^{\rm I}$) is for the DM particles to form orbits completely inside the scatterer (host), and the second ($t_{\rm th}^{\rm II}$) is for attaining kinetic energy equal to the thermal energy of baryons~\cite{acevedo_supernovae_2019}. Thus,
\begin{align}\nonumber
    t_{\rm th}^{\rm WD}=&t_{\rm th}^{\rm I}+t_{\rm th}^{\rm II}\\\simeq & \,10^{-4}\;{\rm yrs}\left(\frac{10^{-40}\ {\rm cm}^2}{\sigma_{\rm n\chi}}\right)\left(\frac{m_\chi}{10^6 \,{\rm GeV}}\right)\left(\frac{1.4 \, M_\odot}{M}\right)^{\frac{3}{2}}\left(\frac{R}{2500\, {\rm km}}\right)^{\frac{7}{2}}\nonumber
    \\
    &+20 \;{\rm yrs}\left(\frac{10^{-40}\ {\rm cm}^2}{\sigma_{\rm n\chi}}\right)\left(\frac{m_\chi}{10^6 \,{\rm GeV}}\right)^2\left(\frac{10^7\,{\rm K}}{\Theta}\right)^{\frac{5}{2}}
    \label{thermWD},
\end{align}
where $\sigma_{\rm n\chi}$ is related to the DM–nucleus cross section ($\sigma_{\rm N\chi}$) via $\sigma_{\rm n\chi} = \sigma_{\rm N\chi} / A^4$, with $A$ being the nuclear mass number (taken as $A=14$ for a carbon–oxygen WD)~\cite{acevedo_supernovae_2019}.
At the end of thermalization, the DM particles settle at the core of the host. EBH can form when the DM core within the thermalization radius satisfies the collapse criterion~\cite{mcdermott_constraints_2012, dasgupta_low_2021}: 
\begin{eqnarray}
 N_X \geq {\rm Max}[N_{\rm Ch}, N_{\rm self}]
 \label{nc}
\end{eqnarray}
where $N_X$ is the total number of  DM particles accumulated within the thermalization radius at the core of the host. $N_{\rm Ch}$ and $N_{\rm self}$ are the Chandrasekhar limit and the number of DM particles required to initiate self-gravitating collapse, respectively~\cite{mcdermott_constraints_2012}. 
If the total number ($N$) of fermions(f)/bosons(b) increases beyond the following limit~\cite{mcdermott_constraints_2012, garani_new_2019}
\begin{eqnarray}
  N^{\rm f}_{\rm Ch}=1.8 \times 10^{51} \left(\frac{100~{\rm GeV}}{m_{\chi}}\right)^3,
  \label{f}
  \end{eqnarray}
or,
\begin{eqnarray}
N^{\rm b}_{\rm Ch} = 1.5 \times 10^{34} \left(\frac{100~{\rm GeV}}{m_{\chi}}\right)^2,
\label{b}
\end{eqnarray}
the gravitational energy dominates over the total thermal energy of DM particles, and gravitational collapse occurs to form a small EBH in the core of a host under the condition of a feeble repulsive DM self-interaction strength~\cite{Dasgupta_2020}. Thus, considering the scenario related to the Chandrasekhar limit (Eq.~\eqref{f} and Eq.~\eqref{b}), bosonic DM particles should achieve the BH formation criterion more easily than fermionic particles. However, this is not true for all cases. The reason is that we also need to consider the effect of $N_{\rm self}$ (see Eq.~\eqref{nc}), as follows: Let us consider, a total DM mass $M_X=N_X m_{\chi}$ is accumulated within a thermal radius of~\cite{mcdermott_constraints_2012}
\begin{eqnarray}
 r_{\rm th}=\left(\frac{9k\T}{4\pi G\rho m_{\chi}}\right)^{1/2}
 \label{rth}
\end{eqnarray}
where the core temperature of the host is $\T$ and $\rho$ is the
baryon density. If the DM density is larger than the
baryon density ($\rho$) within the thermal radius ($r_{\rm th}$)~\cite{mcdermott_constraints_2012} of a host, i.e., 
\begin{eqnarray}
 \frac{3M_X}{4\pi r_{\rm th}^3} \gtrsim \rho ,
 \label{rho}
\end{eqnarray}
the DM particles become self-gravitating.
Substituting Eq.~\eqref{rth} in Eq.~\eqref{rho} we obtain
\begin{eqnarray}
N_X \gtrsim  N_{\rm self} \equiv  \frac{9}{2} \left[\frac{1}{\pi \r m_{\chi}^5}\left(\frac{k\T}{G}\right)^3 \right]^{1/2}.
\label{self}
\end{eqnarray}
This means that the DM becomes self-gravitating once the total number of accumulated DM particles is larger than the critical number $N_{\rm self}$. At this point, one must consider Eqs.~\eqref{f}, \eqref{b}, and \eqref{self} carefully, and use Eq.~\eqref{nc} to proceed further with a specific host.

\begin{figure}
 \centering
 \subfigure[NS]{\includegraphics[width=2.2in]{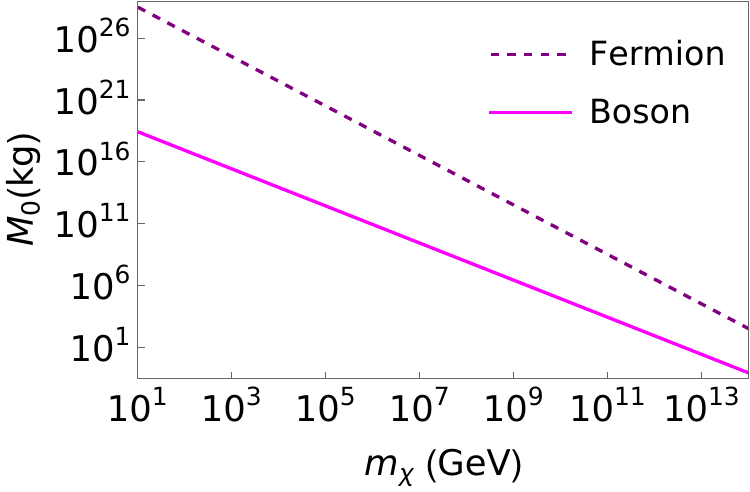}}
\hspace{0.05\textwidth}
 \subfigure[WD]{\includegraphics[width=2.2in]{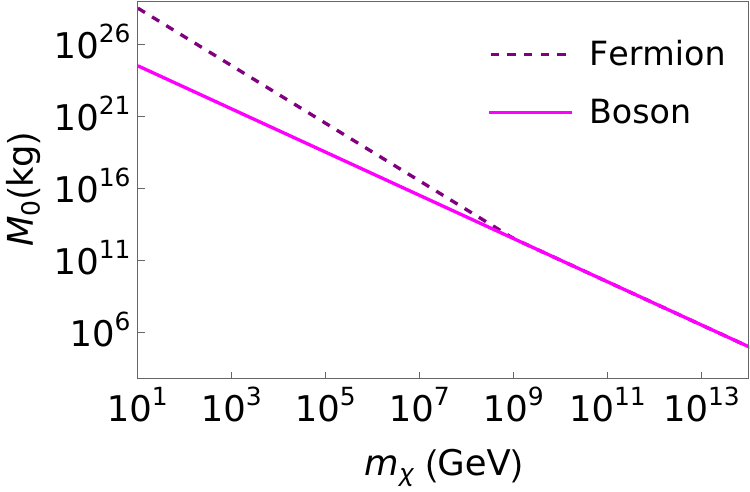}}
 \hspace{0.05\textwidth}
\caption{\label{fm0}Mass ($M_0$) of the endoparasitic collapsed object (BH or naked singularity~\cite{Chakraborty_2024, Chakraborty_low_mass_nakedsingularity2024}) formed inside an NS ($M_{\rm NS}=1.6 M_\odot$, $R_{\rm NS}=12$ km) and a WD ($M_{\rm WD}=0.6 M_\odot$, $R_{\rm WD}=6000$ km) at $t=t_0$ versus the mass ($m_{\chi}$) of DM particles. See section \ref{sec3} for details.}
\end{figure}

If the captured DM particles satisfy the collapse criterion mentioned above, a tiny EBH forms in the core of the host. 
The mass of the newly formed EBH should be $M_0=m_{\chi}{\rm Max}[N_{\rm Ch}, N_{\rm self}] \ll M$, and the formation time ($t_0$) of the said EBH is calculated using Eqs. \eqref{nc}, \eqref{f}, \eqref{b} and \eqref{self} as \cite{dasgupta_low_2021, Chakraborty_2024, Chakraborty_low_mass_nakedsingularity2024, AHA_stall_ang_2025} (and references therein)
\begin{align}
    t_0^{\rm b}=\frac{{\rm Max}[N_{\rm Ch}^{\rm b}, N_{\rm self}]}{F}+t_{\rm th}, \nonumber\\
    t_0^{\rm f}=\frac{{\rm Max}[N_{\rm Ch}^{\rm f}, N_{\rm self}]}{F}+t_{\rm th},
    \label{formation}
\end{align}
due to the accumulation of bosonic and fermionic DM particles, respectively. 
Here, $t_{\rm th}$ is the thermalization time given by Eqs. \eqref{thermNS} and \eqref{thermWD} for NSs and WDs respectively. Panels (a) and (b) of Fig.~\ref{fm0} show the variation of $M_0$ with $m_\chi$ for NSs and WDs.
Note that we are using the multiple-scatter capture rate (Eq.~\eqref{capture}), which can be replaced by the single-scatter capture rate (Eq. (12) of~\cite{mcdermott_constraints_2012}) when one deals with lower mass DM ($m_\chi\lesssim10^6$ GeV). If the collapse criterion is not satisfied, the EBH does not form, and transmutation does not occur. 

\section{\label{sec3.5} Baryon capture rate 
}
Let us consider that a host was born at time $t = 0$, and an EBH of mass $M_0$ is formed at the center of a host by capturing DM particles at time $t=t_0$, as discussed in Sec.~\ref{sec3}. Once an EBH forms, it begins to accrete baryonic matter from its host environment. In most studies~\cite{mcdermott_constraints_2012, dasgupta_low_2021, acevedo_supernovae_2019, Chakraborty_2024}, this process is described using spherical Bondi accretion, sometimes incorporating additional effects such as the rotation of the host system ~\cite{kouvaris_growth_2014, AHA_stall_ang_2025}. The corresponding Bondi accretion rate ($ \dot{m}_{\rm B}$) is given by~\cite{mcdermott_constraints_2012,GenoliniSerpicoTinyakov2021, baumgarte_max_surv_2021, Chakraborty_2024, Chakraborty_low_mass_nakedsingularity2024} 
\begin{equation}
    \dot{m}_{\rm B} \approx \frac{4\pi G^2 m^2 \rho}{c_s^3}
    \label{bondirate}
\end{equation}
where $m$ denotes the instantaneous mass of the EBH at time $t$, $c_s$ is the sound speed of the ambient medium, and the overdot represents differentiation with respect to time.
This description remains valid as long as the accreting medium can be treated within a hydrodynamic framework. 
However, for sufficiently small EBHs, the fluid approximation ceases to be applicable. In particular, the hydrodynamic description breaks down when the de Broglie wavelength of an accreting particle becomes comparable to the horizon size of the EBH itself. In this regime, accretion must be described in terms of individual particle absorption rather than a continuous flow. Accordingly, the mass growth is more appropriately characterized by the absorption cross section ($\sigma_{\rm U}$) derived by Unruh~\cite{Unruh_quantum_1976}. The corresponding quantum accretion rate ($ \dot{m}_{\rm U}$) is given by:~\cite{Griffin_quantum_deathNS}
\begin{equation}
    \dot{m}_{\rm U}=\sigma_{\rm U}\rho v_n
    \label{quantumrate}
\end{equation}
where, 
\begin{align}
    \sigma_{\rm U}=&\frac{2\pi G^2 m^2}{c^4v}\frac{\xi}{1-e^{-\xi}} \nonumber
    \end{align}
with
    \begin{align}
    \xi=&2\pi \frac{G m m_n}{\hbar c} f(v). \nonumber 
\end{align}
Here, $f(v)=(1+v^2)/(v\sqrt{1-v^2})$ and $v=v_n/c$ ($v_n$ is the velocity of the baryon) is the normalized velocity of the baryonic particle of mass $m_n$ being absorbed by EBH.
The accretion rate given by Eq.~\eqref{quantumrate} reduces, in the two asymptotic limits, to

\begin{equation}
\dot{m}_{\rm U} \approx 
\begin{cases}
\text{(i)}\;\;  \frac{2\pi G^2 m^2 \rho}{c^3} & \hspace{1cm} \xi \lesssim 1 ,\\[6pt]
\text{(ii)}\;\;  \frac{2\pi G^2 m^2 \rho}{c^3}\,\xi & \hspace{1cm} \xi \gtrsim 1 
\end{cases}
\label{qrate}
\end{equation}
following~\cite{Unruh_quantum_1976}. 
The two asymptotic expressions in Eq.~\eqref{qrate} correspond to the physically distinct quantum absorption regimes of the EBH. The dimensionless parameter $\xi$ measures the strength of the gravitational interaction between an incoming baryon and the EBH. For $\xi\lesssim1$, the Schwarzschild radius of the EBH is much smaller than the de Broglie wavelength of the incoming baryon, and the absorption probability is quantum-suppressed, leading to the first expression in Eq.~\eqref{qrate}. As the EBH grows, $\xi$ increases. Once $\xi\gtrsim1$, the gravitational interaction becomes sufficiently strong to enhance the absorption probability, yielding the second expression in Eq.~\eqref{qrate}.
The condition $\xi \sim 1$ marks the transition between the two regimes of $\dot{m}_{\rm U}$. The corresponding EBH mass ($M_{\rm U}$) for $\xi \sim 1$ is therefore obtained as:
\begin{equation}
M_{\rm U} \sim \frac{M_{\rm P}^2}{2\pi m_n f(v)},
\label{MU}
\end{equation}
where $M_{\rm P}=\sqrt{\hbar c/G}$ denotes the Planck mass.
Accordingly, expression (i) of Eq.~\eqref{qrate} applies to EBHs that satisfy $M_0 \lesssim M_{\rm U}$, while expression (ii) governs the subsequent evolution at higher masses, up to the onset of Bondi accretion. The latter transition occurs when the EBH horizon has a size comparable to the average interparticle distance in the fluid. The characteristic mass $M_{\rm B}$ that represents this transition is determined by the condition
\begin{equation}
\frac{G M_{\rm B}}{c_{\rm s}^2} \sim \left(\frac{\rho}{m_n}\right)^{-1/3} \hspace{1cm} \Rightarrow \hspace{1cm}  M_{\rm B} \sim \frac{c_{\rm s}^2}{G} \left(\frac{\rho}{m_n}\right)^{-1/3} .
\end{equation}

Thus, three distinct accretion regimes arise,
with the applicable rate determined by the instantaneous mass of EBH $m$ (or, in the beginning, determined by the initial mass $M_0$): Regime (1) for $M_0 \lesssim M_{\rm U}$, the accretion is described by expression (i) of Eq.~\eqref{qrate}; Regime (2) for $M_{\rm U} \lesssim M_0 \lesssim M_{\rm B}$, expression (ii) of Eq.~\eqref{qrate} applies; and Regime (3) for $M_0 \gtrsim M_{\rm B}$, the evolution is governed by the Bondi accretion rate (Eq.~\ref{bondirate}). If an EBH forms with an initial mass $M_0<M_{\rm U}$, it captures baryons at a rate given by (i) of Eq.~\eqref{qrate} and eventually reaches a mass $m>M_{\rm U}$ at which point it starts capturing baryons at a rate given by (ii) in Eq.~\eqref{qrate}. When it grows even further and reaches a mass $m>M_{\rm B}$, it begins to accrete baryons at a rate given by Eq.~\eqref{bondirate}. 
Similarly, an EBH with an initial mass $M_0$ lying between $M_{\rm B}$ and $M_{\rm U}$ starts off by capturing baryons at a rate given by (ii) in Eq.~\eqref{qrate}, eventually reaching the Bondi rate when its mass is above $M_{\rm B}$. An EBH with an initial mass $M_0\gtrsim M_{\rm B}$ starts accreting matter at the Bondi rate and continues until the host is transmuted. However, the growth of an EBH is not guaranteed. Hawking evaporation can dominate over accretion, causing the EBH to lose mass and potentially evaporate completely before reaching the next regime. The actual evolution is therefore determined by the competition between baryonic accretion, dark matter feeding, and Hawking evaporation, as discussed in Sec.~\ref{sec4}.

It is worth noting that in both Regimes (1) and (3), the accretion rate scales quadratically with $m$, that is, $\dot{m} \propto m^2$. In fact, they are related through the following simple relation:
\begin{equation}
  \alpha =  \frac{\dot{m}_{\rm U}|_{\xi\lesssim1}}{\dot{m}_{\rm B}} \approx\frac{1}{2} \left(\frac{c_s}{c}\right)^3.
\end{equation}
In contrast, in Regime (2), the accretion rate exhibits a cubic dependence, that is, $\dot{m} \propto m^3$.

To determine the domain of validity of the three above-mentioned distinct regimes, we consider specific astrophysical systems such as NSs and WDs. 
For an NS, we take $m_n$ to be the neutron mass. The characteristic velocity is set by the Fermi velocity, $v \sim 0.3\text{--}0.4$~\cite{Griffin_quantum_deathNS}, and the typical density is $\rho \approx 1.4 \times 10^{18}\,\mathrm{kg\,m^{-3}}$. These values yield $M_{\rm U} \sim 10^{10}\,\mathrm{kg}$, $M_{\rm B} \approx 8 \times 10^{11}\,\mathrm{kg}$.
For a typical NS, taking $c_s = 0.7c$~\cite{mcdermott_constraints_2012}, one obtains $\alpha \sim 0.2$. This implies that, in Regime (1), the quantum accretion rate is roughly about an order of magnitude smaller than the corresponding Bondi accretion rate.

For a WD, we consider a carbon-oxygen composition with an average nuclear mass number $\sim 14\,\mathrm{u}$. The characteristic velocity is estimated as $v \sim 10^{-3}$ using the thermal velocity $v_{\rm therm} \sim \sqrt{k\Theta/m_n}$ at $\Theta = 10^8\,\mathrm{K}$~\cite{acevedo_supernovae_2019}, and the density is $\rho \sim 10^{9}\,\mathrm{kg\,m^{-3}}$~\cite{acevedo_supernovae_2019}. These parameters give $M_{\rm U} \sim 10^{6}\,\mathrm{kg}$ and $M_{\rm B} \sim 10^{11}\,\mathrm{kg}$. 
For a typical WD, taking $c_s = 0.016c$~\cite{balberg_properties_2000}, we obtain $\alpha \sim 2\times 10^{-6}$. This implies that, in regime (1), the quantum accretion rate is suppressed by approximately six orders of magnitude compared to the Bondi accretion rate.

In the following, we adopt these three regimes of applicability for our analysis.

\section{\label{sec4} Transmutation timescales}
During the accretion phase, the growth and evaporation of the EBH are governed by the following equation:
\begin{subequations}\label{tot}
\begin{numcases}{\dot{m}=}
     k_1 m^2+k_2-\frac{k_3}{m^2} & \text{if $ m \lesssim M_{\rm U} \;\; \text{and} \;\; m \gtrsim M_{\rm B}$} ,\label{tot1}\\[6pt]
     k_1'm^3+k_2-\frac{k_3}{m^2} & \text{if $ M_{\rm U}\lesssim m\lesssim M_{\rm B}$}.\label{tot2}
\end{numcases}
\end{subequations}
The first term on the right-hand side (RHS) of Eq.\eqref{tot1} represents the mass accretion rate ($\dot{m}_{\rm acc}$) of the EBH from its host medium,
\begin{eqnarray}
 \dot{m}_{\rm acc}=k_1 m^2
 \label{acc}
\end{eqnarray}
where $k_1=3\alpha G^2 M/(c_s^3 R^3)$ with $\alpha=1$~\cite{GenoliniSerpicoTinyakov2021, baumgarte_max_surv_2021, Chakraborty_2024, Chakraborty_low_mass_nakedsingularity2024} in the Bondi regime ($m \gtrsim M_{\rm B}$)  and $\alpha=c_{\rm s}^3/2c^3$ in the Unruh $(m \lesssim M_{\rm U})$ regime.
In the intermediate ($M_{\rm U}\lesssim m \lesssim M_{\rm B}$) regime however, Eq.~\eqref{tot2} applies,
where $k_1'=3\pi G^2Mm_n f(v)/( c^3 R^3 M_{\rm P}^2)$.
The third term of the RHS  of Eqs. \eqref{tot1} and \eqref{tot2} accounts for the mass loss rate ($\dot{m}_{\rm Hr}$) due to Hawking radiation, with the evaporation rate given by
\begin{eqnarray}
 \dot{m}_{\rm Hr}=-\frac{k_3}{m^2}
 \label{evp}
\end{eqnarray}
where, $k_3=\hbar c^4/(15360\pi G^2)$~\cite{C_Chakraborty_22_gravmag}. This expression corresponds to the standard Hawking evaporation rate of a non-rotating, uncharged BH.

The second term accounts for mass growth due to the capture of DM particles by EBH. The corresponding capture rate on EBH depends on several factors~\cite{acevedo_supernovae_2019}. As shown in~\cite{janish_type_2019}, DM accretion onto an EBH can proceed through two qualitatively distinct channels, which we briefly discuss below.

A significant DM overdensity may develop in the immediate vicinity of a newly formed EBH, particularly if the collapse of the DM core proceeds with a nonuniform density profile. Within the collisionless spherical approximation~\cite{shapirobook}, a DM population characterized by asymptotic density $\rho_\infty$ and velocity $v_\infty$ far from the EBH accretes at a rate~\cite{janish_type_2019, shapirobook} :
\begin{eqnarray}
 \dot{m}_{\rm capt} = \frac{16\pi \rho_\infty G^2m^2}{v_\infty c^2}.
\end{eqnarray}
This channel of accretion is especially important for bosonic DM, notably when the EBH originates from the collapse of a compact Bose-Einstein condensate (BEC) within a surrounding noncondensed DM core~\cite{janish_type_2019, kouvaris_not-constraining_2013}. Here, we do not consider the possibility of BEC, since the critical temperature for BEC formation is almost always lower than~\cite{dasgupta_low_2021} the core temperature of the stellar objects considered in this work.

In the gravitational potential of the EBH, DM particles undergo repeated scattering with the surrounding stellar medium, gradually losing their energy and angular momentum. Consequently, their orbits shrink inward from the thermalization radius. Once a particle’s angular momentum falls below the critical value, $L < 4 G M_0/c$, it is no longer supported against infall and is promptly captured by the EBH~\cite{janish_type_2019}. Following EBH formation, the system soon achieves a steady state ~\cite{janish_type_2019, acevedo_supernovae_2019} in which the DM supply to the EBH is regulated by~\cite{acevedo_supernovae_2019, janish_type_2019,steigerwald_revisiting_2022}: 
\begin{eqnarray}
  \dot{m}_{\rm DM} =f_{\chi} m_{\chi}F=k_2.
 \label{cap}
\end{eqnarray}
In this work, we assume that DM feeds the EBH efficiently, corresponding to a capture efficiency $f_{\chi}=1$, consistent with the scenarios discussed in~\cite{janish_type_2019} and following the treatment of~\cite{acevedo_supernovae_2019, steigerwald_revisiting_2022}. Accordingly, we adopt Eq.~(\ref{cap}) and substitute it into Eq.~\eqref{tot} as $k_2$ for the analysis presented herein.

In Eq.~\eqref{tot}, the accretion rate $\dot{m}$ may be positive, zero, or negative, corresponding to the growth, stagnation, or evaporation of the EBH, respectively. Physically, the EBH grows by accretion when either of the first two terms or their combined contribution dominates over the third term ($\dot{m}>0$) of Eq.~\eqref{tot}, whereas Hawking evaporation occurs when the third term outweighs the first two terms ($\dot{m}<0$). The condition $\dot{m}=0$ therefore marks the transition between these two regimes.
This transition allows us to define a critical mass, $M_{\rm crit}$, below which the EBH cannot grow and, consequently, no transmutation of the host object can occur. In a realistic and dynamically evolving astrophysical environment, the probability that an EBH attains and sustains the exact mass $M_{\rm crit}$ corresponding to the stagnation condition $\dot{m}=0$ is exceedingly small. Nonetheless, this limiting configuration may be viewed as another manifestation of stalled accretion~\cite{AHA_stall_ang_2025, AHA_2025_MAT}, albeit arising from a different physical mechanism. The expression for the critical mass is obtained by imposing $\dot{m}=0$ in Eq.~\eqref{tot1} and solving the resulting equation for $m$, yielding
\begin{equation}
    M_{\rm crit}=\sqrt{\frac{k_4-k_2}{2k_1}} 
    \label{mcrit}
\end{equation}
where $k_4=\sqrt{k_2^2+4k_1k_3}$. If we approximate both the expressions Eqs. \eqref{tot1} and \eqref{tot2} at the higher $m_\chi$ values by $\dot{m}\approx k_2-k_3/m^2$--since at higher $m_\chi$ the first term is usually negligible compared to the second and third--the approximate part of Eq.~\eqref{mcrit} is applicable to all regimes.
If one ignores the continued DM accretion by the EBH, one obtains
\begin{equation}
    M_{\rm crit1} \equiv M_{\rm crit}|_{k_2\rightarrow0} = \left(\frac{k_3}{k_1}\right)^{1/4}
    \label{mcr}
\end{equation}
in both Regime (3) and Regime (1). If $\dot{m}=0$ occurs in the Regime (2) where Eq.~\eqref{tot2} applies, then
\begin{equation}
    M_{\rm crit1}' = \left(\frac{k_3}{k_1'}\right)^{1/5}.
\end{equation}
However, numerically, $M_{\rm crit1}$ and $M_{\rm crit1}'$ are of the same order of magnitude as each other.

Now, for $\dot{m}>0$, we can numerically integrate Eq.~\eqref{tot} to obtain the transmutation timescale. 
Although there is no closed-form solution that simultaneously accounts for all three regimes with Unruh and Bondi rates (switching between Eq.~\eqref{tot1} and \eqref{tot2}), an analytical solution of Eq.~\eqref{tot1} provides a remarkably accurate result that matches the numerical integrations that consistently incorporate both Eq.~\eqref{tot1} and Eq.~\eqref{tot2} in their respective regimes of validity for NSs and WDs. 
This agreement can be attributed to the dominance of the $k_2$ term over the $k_1' m^3$ contribution in the intermediate regime (Regime 2) and even in Regime (1), rendering the timescale largely insensitive to the baryonic accretion term. Consequently, the closed-form solution of Eq.~\eqref{tot1} obtained below provides an excellent approximation, even when quantum corrections to the accretion process are taken into account. Hence we consider:

\begin{eqnarray}\nonumber
 \int^{t_{\rm d}}_{t_0} dt &=& \int_{M_0}^M \frac{dm}{k_1 m^2+k_2-k_3/m^2}
 \\
\Rightarrow t_{\rm d} &=& t_0 + \frac{1}{k_4\sqrt{2k_1}}. \nonumber
\\
&& \left[\sqrt{k_4+k_2}\tan^{-1}\left\{\frac{\sqrt{2k_1(k_4+k_2)}(M-M_0)}{k_4+k_2+2k_1 M M_0}\right\} \right.
- \left.  \sqrt{k_4-k_2}\tanh^{-1}\left\{\frac{\sqrt{2k_1(k_4-k_2)}(M-M_0)}{k_4-k_2-2k_1 M M_0}\right\} \right]  \nonumber
\\
&\equiv & t_0+t_{\rm acc}
 \label{td}
\end{eqnarray}
where, $t_0$ denotes the EBH formation timescale (see Eq.~\ref{formation}), corresponding to an initial mass $M_0$ and $t_{\rm acc}$ is the characteristic accretion timescale following its formation.
One can rewrite Eq.~\eqref{td} in terms of $M_{\rm crit}$ as 
\begin{align}
    t_{\rm d}=t_0+\frac{M_{\rm crit}}{(k_3+k_1 M_{\rm crit}^4)}
    \left[\sqrt{k_3/k_1}\,\tan^{-1}\left\{\frac{\sqrt{k_3/k_1}\,M_{\rm crit}(M-M_0)}{k_3/k_1+M_{\rm crit}^2MM_0}\right\}-M_{\rm crit}^2\tanh^{-1}\left\{\frac{M_{\rm crit}(M-M_0)}{M_{\rm crit}^2-MM_0}\right\}\right]
    \label{td-mcrit}
\end{align}
which, for  $M \gg M_0$ and $M \gg M_{\rm crit}$, can be approximated as
\begin{align}
    t_{\rm d} \approx t_0+\frac{M_{\rm crit}}{(k_3+k_1 M_{\rm crit}^4)}
    \left[\sqrt{k_3/k_1}\,\tan^{-1}\left\{\frac{\sqrt{k_3/k_1}\,M_{\rm crit}M}{k_3/k_1+M_{\rm crit}^2MM_0}\right\}+M_{\rm crit}^2\tanh^{-1}\left(\frac{M_{\rm crit}}{M_0}\right)\right].
    \label{td-mcrit1}
\end{align}
A close inspection reveals that, for $M_0 \gg M_{\rm crit}$, the first term inside the square bracket of Eq. (\ref{td-mcrit1}) dominates over the second, so that the transmutation timescale $t_{\rm d}$ is effectively governed by the former. In contrast, as $M_0 \rightarrow M_{\rm crit}$, the second term grows rapidly and eventually controls the behavior of $t_{\rm d}$. Indeed, this term diverges at $M_0 = M_{\rm crit}$, implying that the transmutation timescale becomes infinite and that the host object is never transmuted.
This behavior is physically expected: when $M_0 = M_{\rm crit}$, Eq.~(\ref{tot1}) vanishes, yielding a zero accretion rate (see also the discussion above Eq.~\eqref{mcrit}). Consequently, the transmutation timescale diverges. Therefore, Eqs.~(\ref{td}-\ref{td-mcrit1}) are applicable only in the regime $\dot{m}>0$. As $\dot{m}\rightarrow 0$, the second term in the square brackets of Eq.~\eqref{td-mcrit} (or equivalently Eq.~\eqref{td-mcrit1}) diverges. For $\dot{m}\leq 0$, the expression ceases to be valid because the imposed upper integration limit in Eq. (\ref{td}) is no longer satisfied. In this regime, the EBH undergoes evaporation instead, and the associated phenomenology may be of astrophysical interest. We plan to investigate this scenario in a future work~\cite{Adarsha_2026}.

Eq.~(\ref{td-mcrit1}) provides an analytical expression for the transmutation timescale in terms of $k_1$, $k_3$, and $M_{\rm crit}$. 
In the limit $\dot{m}_{\rm capt} \ll \dot{m}_{\rm acc}$ and $\dot{m}_{\rm capt} \ll |\dot{m}_{\rm Hr}|$ -- which effectively corresponds to $k_2 \rightarrow 0$ -- one may simply replace $M_{\rm crit}$ by $M_{\rm crit1}$ from Eq.~\eqref{mcr} in Eq.~\eqref{td-mcrit} to obtain the corresponding transmutation timescale (see Eq.~\eqref{tdk20}) for the scenario in which DM accretion onto the EBH is inefficient. Note that if EBH evaporation and DM capture rates during the accretion phase are negligibly low compared to the Bondi accretion rate 
(i.e., $\dot{m}_{\rm capt} \ll \dot{m}_{\rm acc}$ and $|\dot{m}_{\rm Hr}| \ll \dot{m}_{\rm acc}$), one readily recovers the previously obtained 
accretion timescale ($t_{\rm acc}^{\rm earlier}$),
\begin{eqnarray}
t_{\rm acc}^{\rm earlier}
= \frac{c_s^3 R^3}{3 G^2 M M_0}
\equiv \frac{1}{k_1 M_0},
\label{tacc}
\end{eqnarray}
from Eq.~\eqref{td} in the limit $M \gg M_0$ (see Sec.~\ref{ss1}). 
This expression agrees with earlier results in the literature 
(see, e.g., Eq.~(5.3) of~\cite{Chakraborty_low_mass_nakedsingularity2024}, Eq.~(48) of~\cite{GenoliniSerpicoTinyakov2021}, 
Eq.~(37) of~\cite{mcdermott_constraints_2012},~\cite{baumgarte_max_surv_2021} and references therein).
In this limit, the corresponding transmutation timescale ($t_{\rm d}^{\rm earlier}$) becomes
\begin{equation}
t_{\rm d}^{\rm earlier} = t_0 + t_{\rm acc}^{\rm earlier}
\label{earlier}
\end{equation}
 (see Eq.~(5.3) of~\cite{Chakraborty_low_mass_nakedsingularity2024}).
However, to determine the actual time $t_{\rm d}$ required for a host object to evolve
from its formation to complete transmutation into a BH, one must retain
the full expression given in Eq.~\eqref{td-mcrit}.

\section{\label{sec5}Results: Transmutation timescales for different astrophysical objects}

In this section, we apply the above formulation to MSPs and WDs. MSPs are usually older among the pulsars, because they are spun up by accreting matter from a companion star in a binary system. While the oldest MSPs have been estimated to be as old as $\sim10$ Gyr, such estimates are subject to considerable uncertainty and may be overestimated~\cite{Kiziltan_2010}. Therefore, we adopt a more conservative characteristic age of 1 Gyr for our constraints.
WDs, on the other hand, can have a wide age range. Older WDs are generally cooler and carry small magnetic fields. In this section, we consider less magnetized MSPs (we will ignore the hypothetical millisecond magnetars) and slowly rotating (spin period $\gtrsim10^4$ s), less magnetized, ordinary WDs. 
We will consider MSPs and WDs located in three distinct regions: the Galactic disk, Galactic bulge, and globular clusters (GCs), since the density ($\rho_{\rm DM}$) of DM depends on the location of the host. 
For instance, the average value of $\rho_{\rm DM}$ varies for the Galactic disk, Galactic bulge, and GCs as follows: $\rho_{\rm disk}=0.4$ GeV.cm$^{-3}$~\cite{Hooper_2017}, $\rho_{\rm bulge}=10^3-7 \times 10^4$ GeV.cm$^{-3}$~\cite{bramante_detecting_2014, Hooper_2017} and $\rho_{\rm GC}=10^3$ GeV.cm$^{-3}$~\cite{mccullough_capture_2010}. \footnote{The DM density in GCs is debated, and the observations can be explained by very low ($\lesssim40$ GeV.cm$^{-3}$)~\cite{Hooper_2017} up to a few $\times 10^4$ GeV.cm$^{-3}$~\cite{Garani_2023_GC_density}. 
Here, we take a value that is an order of magnitude lower than the upper limit, since it is not outright ruled out by observational constraints and has been previously used in the literature.} 
We estimate the growth rate of EBH ($\dot{m}$) and the transmutation timescale ($t_{\rm d}$) using Eqs. \eqref{td} and \eqref{earlier}, and constrain the $m_\chi-\sigma_{\rm n\chi}$ space for bosonic and fermionic DM from the MSP and WD age at the above mentioned locations.

\subsection{\label{NS} Millisecond pulsars}

\begin{figure}[]
    \centering
\subfigure[For Galactic disk ($\rho_{\rm disk}=0.4$ GeV.cm$^{-3}$)]{\includegraphics[width=2.5in]{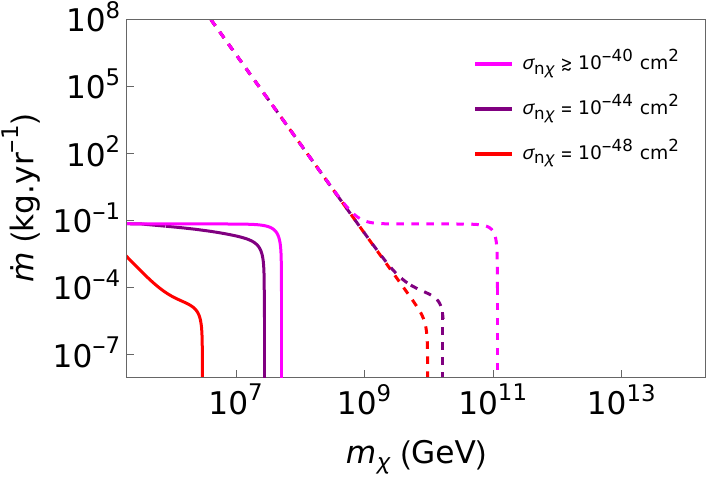}}
\hspace{0.05\textwidth}
\subfigure[For Galactic bulge ($\rho_{\rm bulge}=7\times10^4$ GeV.cm$^{-3}$)]{\includegraphics[width=2.5in]{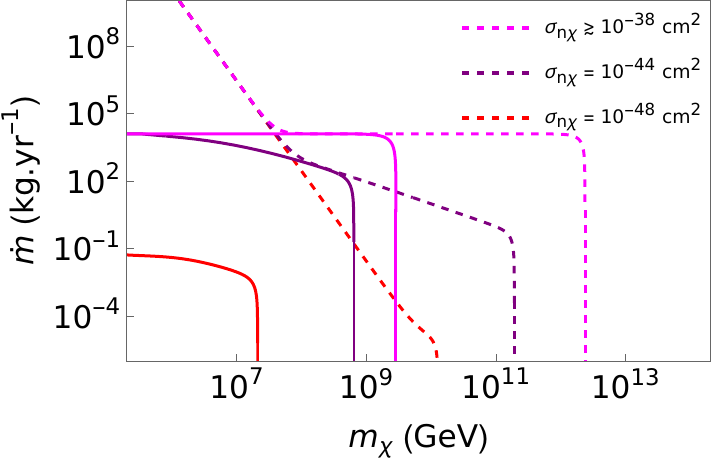}}
\hspace{0.05\textwidth}
\subfigure[For Globular cluster ($\rho_{\rm GC}=10^3$ GeV.cm$^{-3}$)]{\includegraphics[width=2.5in]{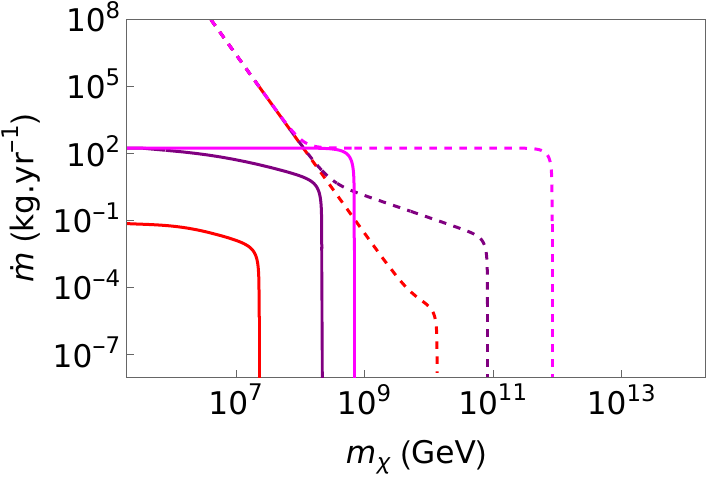}}
\hspace{0.05\textwidth}
\subfigure[$\dot{m}$ considered to be independent of $\rho_{\rm DM}$ and $\sigma_{\rm n\chi}$, i.e., $k_2 \rightarrow 0$]{\includegraphics[width=2.5in]{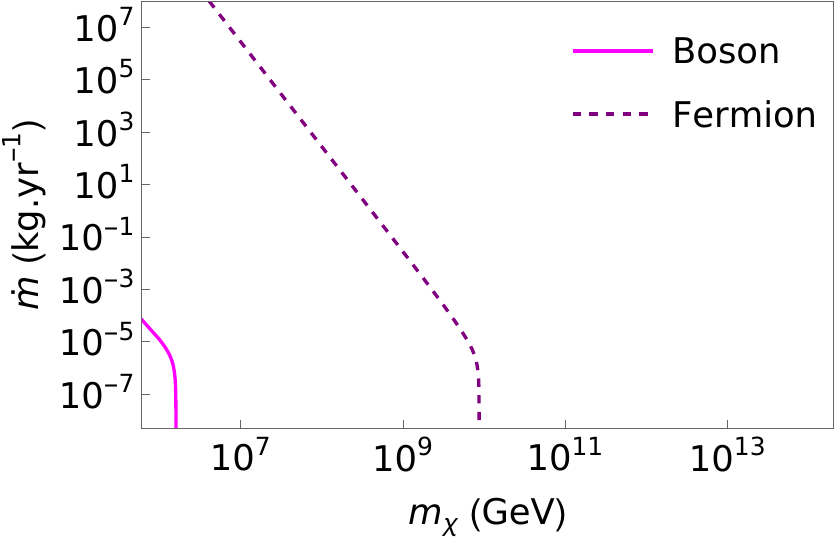}}

\caption{Growth rate ($\dot{m}$) of an EBH when it forms at the core of an MSP of mass $M \sim 1.6M_{\odot}$ and $R \sim 12$ km as a function of the DM particle mass $m_\chi$ for different values of $\rho_{\rm DM}$ and $\sigma_{\rm n\chi}$. Here, the solid (dashed) curves correspond to bosonic (fermionic) DM. Panels (a)–(c) demonstrate that, for a fixed $\rho_{\rm DM}$, decreasing the scattering cross-section from $\sigma_{\rm n \chi} \gg \sigma_{\rm crit}^{\rm MSP}$ to $\sigma_{\rm n\chi} \ll \sigma_{\rm crit}^{\rm MSP}$, where $\sigma_{\rm crit}^{\rm MSP} \sim 10^{-45}{\rm cm}^2$, causes the growth rate $\dot{m}$ to approach the $k_2\rightarrow0$ limit, as illustrated in panel (d). In contrast, for $\sigma_{\rm n\chi} \gg \sigma_{\rm crit}^{\rm MSP}$, panels (a)–(c) show that $\dot{m}$ becomes nearly independent of $m_\chi$ over a certain range of DM masses for both bosonic and fermionic DM. At low $m_\chi$, the growth is dominated by Bondi accretion (visible only for fermions in panels (a), (b), and (c)), whereas at high $m_\chi$ the effect of Hawking radiation becomes dominant, leading to a steep decline in $\dot{m}$. The $m_\chi$-independent plateau observed in panels (a), (b), and (c) arises from the dominance of the DM capture term ($k_2$ in Eq.~\eqref{tot}). Panel (d) corresponds to the case where DM capture is neglected, retaining only the $k_1$ and $k_3$ terms in Eq.~\eqref{tot}. This panel is included for comparison with panels (a)–(c) to highlight the crucial role played by the capture term. Further details are discussed in Sec.~\ref{NS}.}
    \label{MdotNS}
\end{figure}

The transmutation timescale $t_{\rm d}$ for MSPs can be calculated once we know the domain of validity of Eq.~\eqref{td}. 
Let us consider an MSP of mass $M \sim 1.6M_{\odot}$ and $R \sim 12$ km (corresponding to $\sigma_{\rm crit}^{\rm MSP}\approx10^{-45}{\rm cm}^2$), where $M_{\odot}$ is the solar mass unit. To determine the domain, we check the growth rate of the EBH immediately after its formation against $m_\chi$ using Eq.~\eqref{tot}, as shown in panels (a), (b), and (c) of Fig.~\ref{MdotNS} corresponding to different $\rho_{\rm DM}$ and different $\sigma_{\rm n\chi}$. 
Panels (a), (b) and (c) show that the domain of validity changes for different $\sigma_{\rm n\chi}$ for a fixed $\rho_{\rm DM}$.
Contrary to the simple inverse-square drop of $\dot{m}$ expected in the case where only Bondi accretion is considered, a complex interplay between the baryon accretion rate, DM capture rate, and Hawking radiation rate is observed to influence the growth of EBH. The DM capture rate and Hawking radiation terms (second and third terms in Eq.~\eqref{tot1}) have a huge impact at higher $m_\chi$ values for $\sigma_{\rm n\chi}\gtrsim\sigma_{\rm crit}^{\rm MSP}$, where the initial mass of the EBH is smaller. 
However, for $\sigma_{\rm n\chi}\ll \sigma_{\rm crit}^{\rm MSP}$, capture rate is negligible, making $k_2$ extremely small; hence, the growth rate resembles that shown in panel (d). This explains why an EBH formed with an initial mass $M_0$ in Regimes (1) or (2) (see Sec.~\ref{sec3.5}) still provides a transmutation timescale that closely agrees with the solution given by Eq.~\eqref{td}. If the growth timescale is dominated by the initial growth rate and the initial growth rate is determined by the nearly $m_\chi$-independent $k_2$ term (i.e., the DM capture rate for $\sigma_{\rm n\chi}\gg \sigma_{\rm crit}^{\rm MSP}$), then the results of numerical calculations must agree with the results of Eq.~\eqref{td}. 
 Even in cases where $k_2$ is not dominant (as in, for example, the $\sigma_{\rm n\chi}=10^{-48}$ cm$^2$ red curves in panels (a), (b) and (c) of Fig.~\ref{MdotNS}) one finds that the effect of quantum accretion rates is small enough so that the results of Eq.~\eqref{td} agree with the numerical calculations.
A careful inspection reveals that the Hawking radiation term steepens the $\dot{m}$ curve, whereas the DM capture is responsible for the plateau, where, regardless of the initial mass of the EBH, the growth rate is fixed.
Since at higher $m_\chi$ values, $M_0$ is smaller, the capture of DM particles adds significant mass to the EBH sufficiently fast to keep the growth rate constant for a range of $M_0$ if $\sigma_{\rm n\chi}\gtrsim\sigma_{\rm crit}^{\rm MSP}$. This contrasts starkly with even considering only the effects of baryon accretion and Hawking radiation as depicted in panel (d) of Fig.~\ref{MdotNS}, which is equivalent to $\sigma_{\rm n\chi}\ll\sigma_{\rm crit}^{\rm MSP}$.
From the figure, we notice that if one were to consider only the host mass accretion rate and Hawking radiation (panel (d)), one would conclude that the EBH made of bosonic (fermionic) DM particles above $m_\chi\sim10^6$ GeV ($m_\chi\sim10^{10}$ GeV) would quickly evaporate and disappear. The minimum EBH mass required for growth in this scenario is given by Eq.~\eqref{mcr}, yielding $M_{\rm crit1}\sim10^{10}$ kg.

When the full growth equation, Eq.~\eqref{tot1} is considered, these limits shift to $m_\chi \sim 10^7$ GeV for bosonic DM and $m_\chi \sim 10^{11}$ GeV for fermionic DM, both corresponding to a critical EBH mass $M_{\rm crit}\sim10^8$ kg obtained from Eq.~\eqref{mcrit} for $\sigma_{\rm n\chi}\gtrsim10^{-40}$ cm$^2$ and $\sigma_{\rm n\chi}\gtrsim10^{-38}$ cm$^2$ (for bosons and fermions, respectively) in the Galactic disk environment (panel (a) of Fig.~\ref{MdotNS}). For the Galactic bulge and the GC, the corresponding critical masses are $M_{\rm crit}\approx6\times10^5$ kg and $\approx5\times10^6$ kg, respectively, with $\dot{m}\to0$ at the $m_\chi$ values associated with these critical masses, as shown in panels (b) and (c). Note that the lowest $M_{\rm crit}$ corresponds to $\sigma_{\rm n\chi}\gtrsim10^{-36}$ cm$^2$ where DM capture rate saturates and hence $k_2$ becomes independent of $\sigma_{\rm n\chi}$.

\begin{figure}
    \centering
    \includegraphics[width=2.5in]{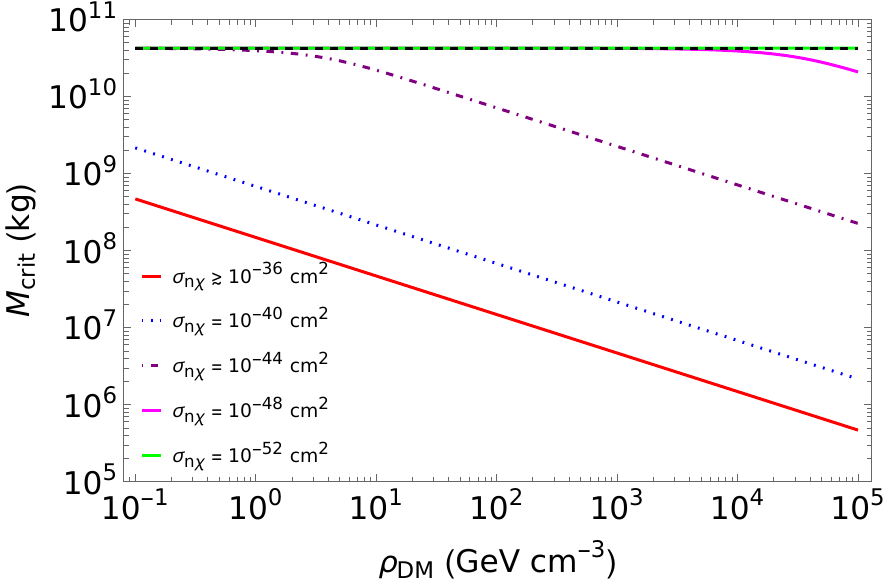}
    \caption{$M_{\rm crit}$, given by Eq.~\eqref{mcrit} inside an MSP as a function of the DM density $\rho_{\rm DM}$ in its surroundings and $\sigma_{\rm n\chi}$. The black dashed line indicates the $M_{\rm crit1}$ given by Eq.~\eqref{mcr}, which is a constant.  For details refer to Sec.~\ref{NS}.}
    \label{Mcrit-MSP}
\end{figure}

In fact, one can plot the minimum mass, $M_{\rm crit}$, of an EBH that can form in MSPs residing in environments with different DM densities  and for different $\sigma_{\rm n\chi}$, as shown in Fig.~\ref{Mcrit-MSP}. While $M_{\rm crit}$ depends on $\rho_{\rm DM}$, the mass $M_{\rm crit1}$ is independent of $\rho_{\rm DM}$ and hence appears as a constant, shown by the black dashed line in the figure. 
As $\sigma_{\rm n\chi}$ decreases, $M_{\rm crit}$ almost coincides with the black dashed line for lower $\rho_{\rm DM}$ since for $\sigma_{\rm n\chi}\ll\sigma_{\rm crit}^{\rm MSP}$, effectively, $k_2$ vanishes.
From Fig.~\ref{Mcrit-MSP}, one can infer that as one moves away from DM-dense regions, an EBH must be more massive at birth in order to grow and eventually transmute its host MSP. This, in turn, implies that the DM particles must be lighter (see Fig.~\ref{fm0}) if they are to transmute MSPs located in low DM-density environments. For instance, in the Galactic disk, where the DM density is
$\rho_{\rm disk}=0.4$ GeV cm$^{-3}$~\cite{Chakraborty_2024, Chakraborty_low_mass_nakedsingularity2024}, bosonic (fermionic) DM particles must satisfy $m_\chi \lesssim 10^7$ GeV ($m_\chi \lesssim 10^{11}$ GeV) to form an EBH of at least the minimum mass required for sustained growth ($\dot{m}>0$) and subsequent transmutation of the host MSP into a black hole, provided $\sigma_{\rm n\chi}\gg\sigma_{\rm crit}^{\rm MSP}$. For lower $\sigma_{\rm n\chi}$ the DM particles must be lighter still to attain the minimum $M_0$ for EBH to grow.
\\

Before calculating $t_{\rm d}$ for MSPs, we must first check which specific parameters (i.e., $N_{\rm Ch}^{\rm b}, N_{\rm Ch}^{\rm f}, N_{\rm self}$) are involved in the calculation of transmutation timescales. 
Substituting $\rho \approx 1.4 \times 10^{18}$ kg.m$^{-3}$~\cite{mcdermott_constraints_2012} in Eq.~\eqref{self}, we obtain the following expression for the asymmetric DM particles required to initiate the self-gravitating collapse in a typical NS:
\begin{eqnarray}
 N_{\rm self}^{\rm NS}=4.8 \times 10^{41} \left(\frac{100~{\rm GeV}}{m_{\chi}}\right)^{5/2}\left(\frac{\Theta}{10^5~{\rm K}}\right)^{3/2}
 \label{nselfNS}
\end{eqnarray}
where, the core temperature $\Theta$ is taken as $\sim 10^5$ K~\cite{mcdermott_constraints_2012}. Primarily, we confine our study to DM particles in the mass range $m_{\chi}\sim10^6-10^{14}$ GeV, since multiple scattering is necessary only above $\sim10^6$ GeV~\cite{bramante_multiscatter_2017, acevedo_supernovae_2019}. We can now obtain the formation timescale for EBH inside NSs, for the above mentioned DM mass range, by replacing $N_{\rm self}$ in Eq.~\eqref{self} by $N_{\rm self}^{\rm NS}$ in Eq.~\eqref{nselfNS}. 

We now apply Eq.~\eqref{td} to compute the numerical value of the transmutation timescale, $t_{\rm d}$, for an MSP capturing DM particles from its surroundings. The solution is evaluated within the domain of validity of Eq.~\eqref{td} as discussed in Sec.~\ref{sec4}, which requires $M_0 \gtrsim M_{\rm crit}$. Consequently, the calculation naturally terminates at the value of $m_\chi$ for which $M_0 \sim M_{\rm crit}$, beyond which Eq.~\eqref{td} ceases to apply.
At exactly the $m_\chi$ for which $M_0=M_{\rm crit}$ (for both bosonic and fermionic DM), the analytic structure of Eq.~\eqref{td} implies $t_{\rm d} \rightarrow \infty$. Although this divergence arises from the logarithmic divergence of the $\tanh^{-1}$ term and is therefore not explicitly resolved on the scale of the figure, its presence is manifested as a vertical asymptote in the solution. The apparent termination of the $t_{\rm d}$ curves thus reflects the intrinsic divergence of the timescale at $M_0 = M_{\rm crit}$.
Using the above parameters, for an MSP of mass $1.6\,M_\odot$ and radius $12~\mathrm{km}$, we compute the DM capture rate $F$ for various values of the scattering cross-section $\sigma_{\rm n\chi}$ and the DM particle mass $m_\chi$ from Eq.~\eqref{capture}. Subsequently, we evaluate the transmutation timescale $t_{\rm d}$ for fermionic and bosonic DM using Eq.~\eqref{formation}. The parameters $\sigma_{\rm n\chi}$ and $m_\chi$ are treated as free parameters, which can be constrained by requiring that the transmutation timescale satisfy $t_{\rm d} > 1~\mathrm{Gyr}$, motivated by the fact that the average age of MSPs is $\sim 1~\mathrm{Gyr}$~\cite{Kiziltan_2010}.

\begin{figure}[h!]
 \centering
\subfigure[For Galactic disk ($\rho_{\rm disk}=0.4$ GeV.cm$^{-3}$)]{\includegraphics[width=2.1in]{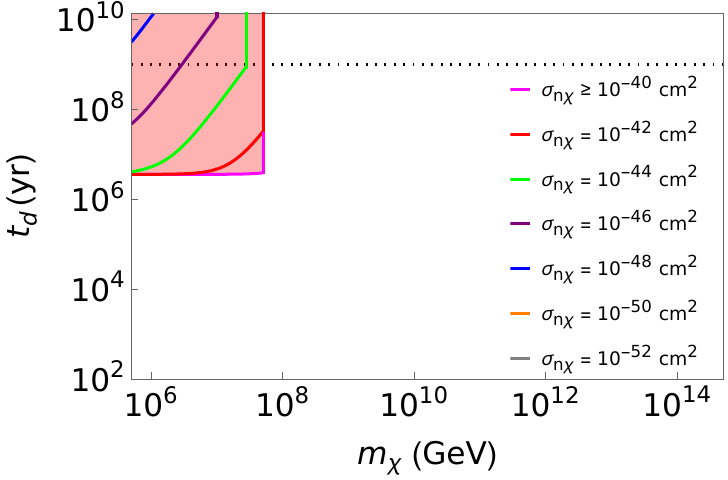}}
\hspace{0.01\textwidth}
\subfigure[For Galactic bulge ($\rho_{\rm bulge}=7 \times 10^4$ GeV.cm$^{-3}$)]{\includegraphics[width=2.1in]{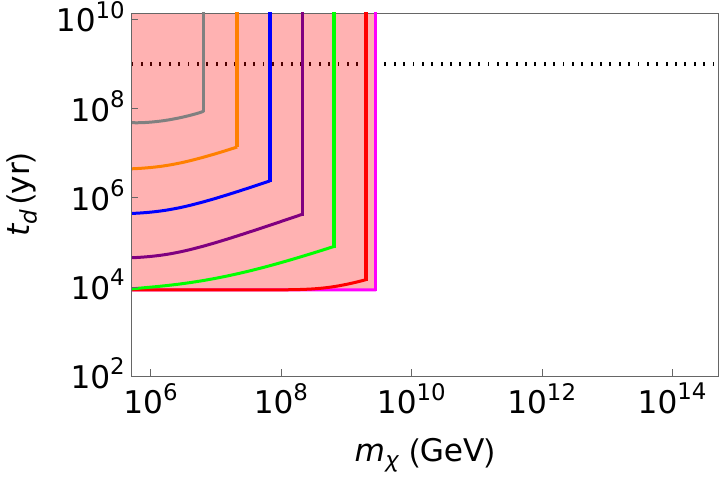}}
\hspace{0.01\textwidth}
\subfigure[For globular cluster ($\rho_{\rm GC}=10^3$ GeV.cm$^{-3}$)]{\includegraphics[width=2.1in]{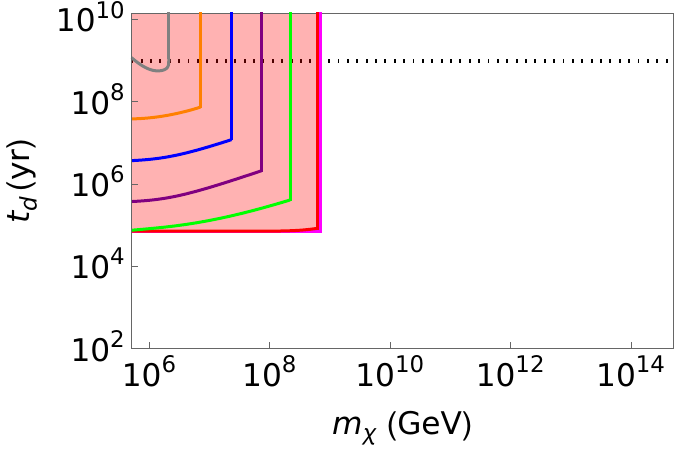}}
\hspace{0.01\textwidth}
\subfigure[For Galactic disk ($\rho_{\rm disk}=0.4$ GeV.cm$^{-3}$)]{\includegraphics[width=2.1in]{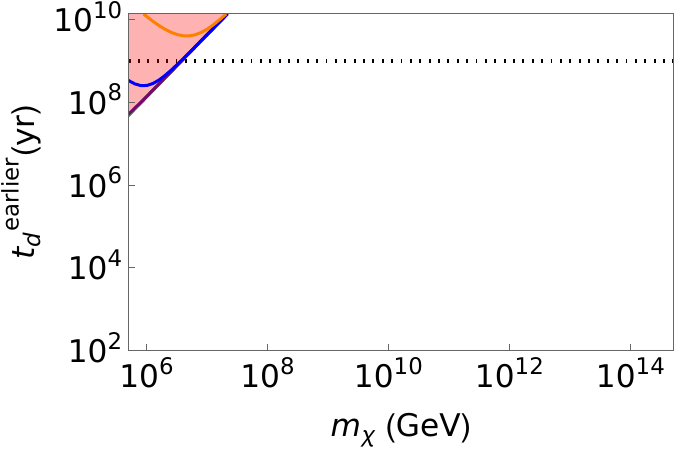}}
\hspace{0.01\textwidth}
\subfigure[For Galactic bulge ($\rho_{\rm bulge}=7 \times 10^4$ GeV.cm$^{-3}$)]{\includegraphics[width=2.1in]{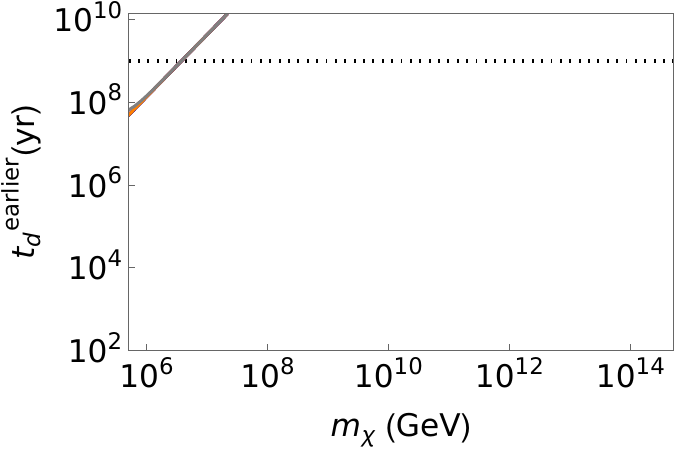}}
\hspace{0.01\textwidth}
\subfigure[For globular cluster ($\rho_{\rm GC}=10^3$ GeV.cm$^{-3}$)]{\includegraphics[width=2.1in]{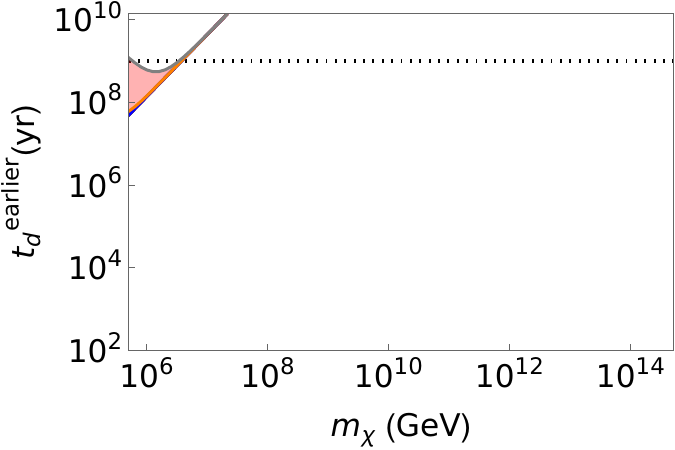}}
\hspace{0.01\textwidth}

\caption{\label{NSboson} 
Transmutation time $t_{\rm d}$ as a function of the DM particle mass $m_\chi$ for a millisecond pulsar (MSP) capturing asymmetric bosonic DM particles, located in the Galactic disk (Panels (a) and (d)), Galactic bulge (Panels (b) and (e)), and globular clusters (Panels (c) and (f)). The shaded regions correspond to DM–nucleon scattering cross-sections in the range $\sigma_{\rm n\chi} \in [10^{-52},10^{-40}] {\rm cm}^2$. Panels (a)–(c) are obtained using Eq.~\eqref{td}, whereas panels (d)–(f) are computed using Eq.~\eqref{earlier}. We assume that the MSP captures asymmetric bosonic DM particles with a velocity dispersion $\bar v = 220~{\rm km~s^{-1}}$, and the sound speed inside the MSP is taken to be $c_s = 0.7c$~\cite{mcdermott_constraints_2012}. The black dotted line denotes the typical age of MSPs, $T_{\rm MSP} \sim 1~{\rm Gyr}$~\cite{Kiziltan_2010}. The apparent cutoff at certain values of $m_\chi$ in panels (a)–(c) arises due to the divergence $t_{\rm d} \rightarrow \infty$ when the initial mass satisfies $M_0 = M_{\rm crit}$, corresponding to $\dot m \rightarrow 0$. Significant differences are evident between the transmutation times $t_{\rm d}$ and $t_{\rm d}^{\rm earlier}$. For further discussion, see Sec.~\ref{NS}.}

\end{figure}

First, we consider an MSP located in the galactic disk. Panel (a) of Fig.~\ref{NSboson} shows the transmutation timescale ($t_{\rm d}=t_0+t_{\rm acc}$) using Eq.~\eqref{td}, plotted against the mass of bosonic DM particles ($m_\chi$) for $\sigma_{\rm n\chi} \in [10^{-52},10^{-40}]$ cm$^2$. For $\sigma_{\rm n\chi}>10^{-40}$ cm$^2$, the timescale is saturated and coincides with the $\sigma_{\rm n\chi}=10^{-40}$ cm$^2$ curve and hence the legend shows $\sigma_{\rm n\chi}\geq10^{-40}$ cm$^2$ for the magenta curve. One notices a clear departure from the earlier timescale calculation (panel (d)), which calculates $t_{\rm d}^{\rm earlier}$ (Eq.~\eqref{earlier}) by accounting for only the accretion timescale once the EBH is formed. 
A similar conclusion holds for regions with higher DM densities, such as the Galactic bulge and GC. For concreteness, we adopt $\rho_{\rm bulge}=7 \times 10^4$ GeV.cm$^{-3}$ ~\cite{Chakraborty_2024, Chakraborty_low_mass_nakedsingularity2024} and $\rho_{\rm GC}=10^3$ GeV.cm$^{-3}$. We find that, over the entire range $\sigma_{\rm n\chi} \in [10^{-52},10^{-40}]$ cm$^2$, the transmutation timescale $t_{\rm d}$ computed using the revised formulation deviates significantly from that obtained using the earlier prescription. Moreover, owing to the enhanced DM density, the resulting $t_{\rm d}$ can be reduced by several orders of magnitude compared to the corresponding values in lower-density environments.
Another peculiar trend visible in the case of bosonic DM particles is that there is a certain range of $m_\chi$ and $\sigma_{\rm n\chi}$ for which $t_{\rm d}$ is almost independent of the initial mass of the EBH ($M_0$) or, in turn, independent of $m_\chi$. This is almost coincident with the plateau region in $\dot{m}$ (Fig.~\ref{MdotNS}), but only for larger $\sigma_{\rm n\chi}$.  Because the formation time $t_0$ decreases with $m_\chi$ (refer to Fig.~\ref{timescale-split-NS} below), the total timescale is dominated by $t_{\rm acc}$ in Eq.~\eqref{td} for $\sigma_{\rm n\chi}\sim10^{-50}$ cm$^2$ and above. However, for $10^{-52}$ cm$^2$, as seen in panel (c) of Fig.~\ref{NSboson} (gray curve), $t_0$ becomes comparable to or greater than $t_{\rm acc}$, because the decreasing $t_{\rm d}$ that we see near $10^6$ GeV is due to $t_0$ and the cutoff comes from diverging $t_{\rm acc}$ (refer to panel (a) of Fig.~\ref{timescale-split-NS} below to see a similar trend in the case of Galactic disk at $\sigma_{\rm n\chi}=10^{-50}$ cm$^2$). 
The plateau in $t_{\rm d}$ curves corresponding to $\sigma_{\rm n\chi}=10^{-40}$ cm$^2$ (i.e., $\sigma_{\rm n\chi}\gg\sigma_{\rm crit}^{\rm MSP}$) observed in panels (a), (b), and (c) of Fig.~\ref{NSboson} are due to the feature of the DM accretion rate given by $k_2$ in Eq.~\eqref{tot}, which becomes independent of $\sigma_{\rm n\chi}$ and $m_\chi$. This is further discussed in Appendix \ref{ss2}. For lower $\sigma_{\rm n\chi}$ however, this feature of $k_2$ term is no longer observed as it becomes dependent on $m_\chi$ as well as the fact that it competes with the Hawking evaporation term ($k_3/m^2$ in Eq.~\eqref{tot}). Panels (d), (e) and (f) plot $t_{\rm d}^{\rm earlier}$ (Eq.~\eqref{earlier}), which shows mostly the Bondi rate-dictated rapid rise of $t_{\rm d}^{\rm earlier}$ with $m_\chi$ in panels (e) and (f). In panel (d), the part of the curve determined by the decreasing formation time can be observed. Since in $t_{\rm d}^{\rm earlier}$, growth time is independent of $\sigma_{\rm n\chi}$, all the different curves corresponding to $\sigma_{\rm n\chi}$ coincide which is seen clearly in panels (e) and (f).

\begin{figure}[h!]
 \centering
 \subfigure[For Galactic disk ($\rho_{\rm disk}=0.4$ GeV.cm$^{-3}$)]{\includegraphics[width=2.1in]{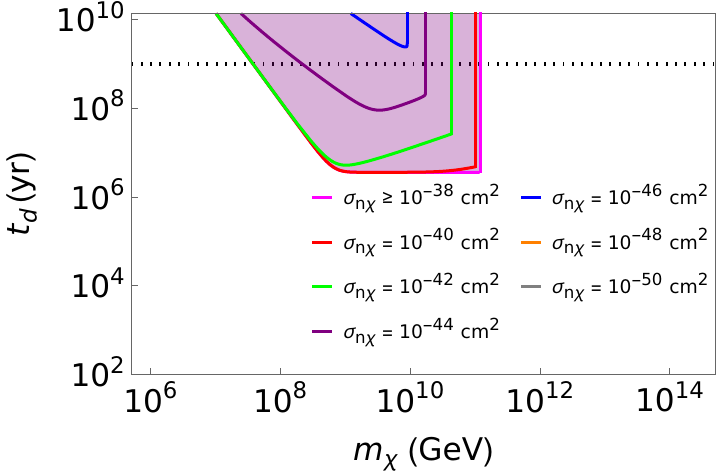}}
\hspace{0.01\textwidth}
\subfigure[For Galactic bulge ($\rho_{\rm bulge}=7 \times 10^4$ GeV.cm$^{-3}$)]{\includegraphics[width=2.1in]{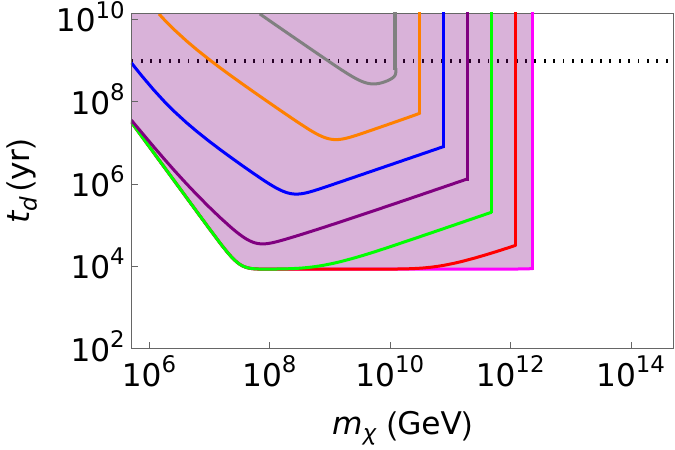}}
\hspace{0.01\textwidth}
\subfigure[For globular cluster ($\rho_{\rm GC}=10^3$ GeV.cm$^{-3}$)]{\includegraphics[width=2.1in]{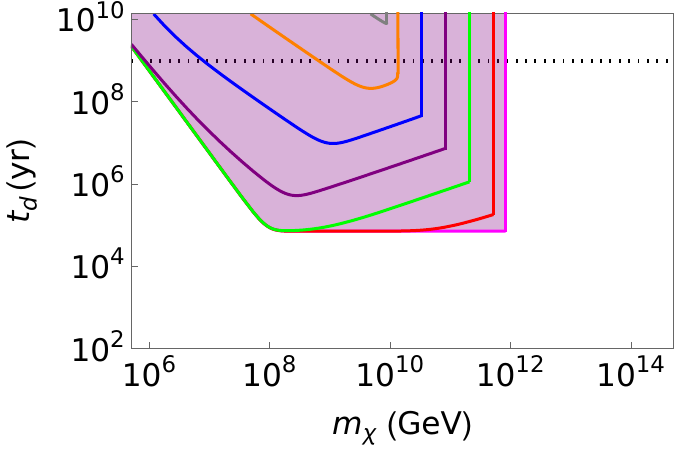}}
\hspace{0.01\textwidth}
\subfigure[For Galactic disk ($\rho_{\rm disk}=0.4$ GeV.cm$^{-3}$)]{\includegraphics[width=2.1in]{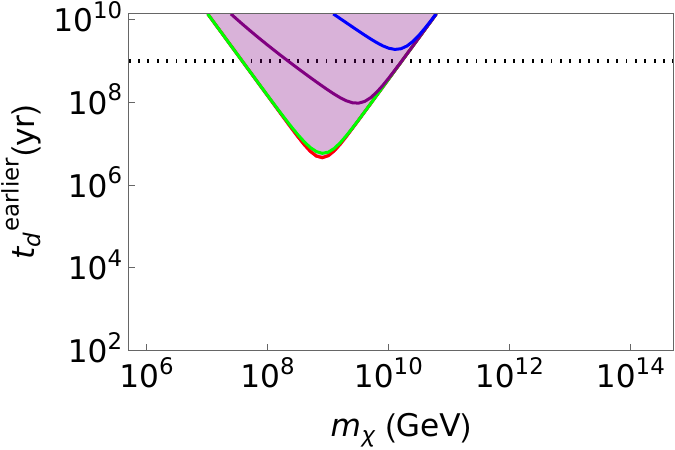}}
\hspace{0.01\textwidth}
\subfigure[For Galactic bulge ($\rho_{\rm bulge}=7 \times 10^4$ GeV.cm$^{-3}$)]{\includegraphics[width=2.1in]{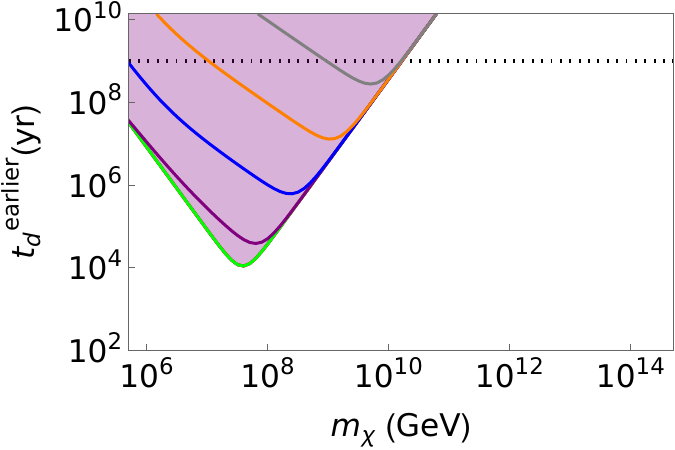}}
\hspace{0.01\textwidth}
\subfigure[For globular cluster ($\rho_{\rm GC}=10^3$ GeV.cm$^{-3}$)]{\includegraphics[width=2.1in]{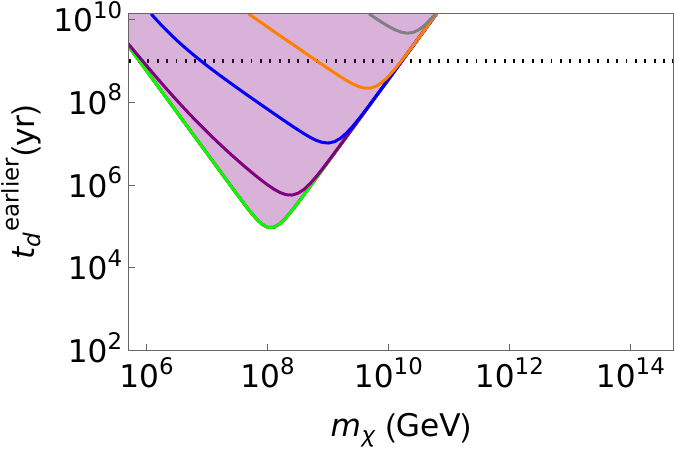}}

\caption{\label{NSfermion}Same as Fig.~\ref{NSboson} but for asymmetric fermionic DM captured by MSPs. A noticeable difference is that the panels (a)–(c) resemble (d)–(f) respectively towards the lower-edge of $m_{\chi}$, unlike in the case of bosonic DM (Fig.~\ref{NSboson}). For further details, refer to section \ref{NS}.}

\end{figure}

We can plot the same for fermionic DM particles, as shown in Fig.~\ref{NSfermion}. In case of fermionic DM, we see that $t_{\rm d}$ drops as $m_{\chi}$ increases in the lower range of $m_\chi$ and then attains constancy (plateau) or rises again.
The kind of $m_\chi$-independent behavior of $t_{\rm d}$ we observe for a wide range of $m_\chi$ in the case of bosonic DM for higher $\sigma_{\rm n\chi}$ is not observed for such a wide range of $m_\chi$ in the case of fermionic DM. In panels (a), (b), and (c), we see that a plateau of a similar kind appears only for $\sigma_{\rm n\chi}\geq 10^{-40}$ cm$^2$ in the case of fermionic DM. 
Similar to the case of bosonic DM, for lower $\sigma_{\rm n\chi}$, while the total timescale is still dominated by $t_{\rm acc}$ at higher $m_\chi$, the DM accretion term $k_2$ is not $m_\chi$-independent and competes with Hawking evaporation; hence, we obtain slanted flat regions. 
Panels (d), (e), and (f) plot $t_{\rm d}^{\rm earlier}$, which resemble the $t_{\rm d}$ plots closely in the lower $m_\chi$ range, where $t_0$ dominates. 
The difference is due to the growth time which in case of $t_{\rm d}^{\rm earlier}$ is dictated by only Bondi accretion rate which decreases with $m_\chi$ and hence the rapid increase in $t_{\rm d}^{\rm earlier}$ rather than the extended plateau region and then cutoff observed in panels (a), (b) and (c).

 Fig.~\ref{timescale-split-NS} shows the formation timescale ($t_0$) and the accretion timescales plotted separately, taking $\rho_{\rm disk}=0.4$ GeV.cm$^{-3}$ and different $\sigma_{\rm n\chi}$ values for bosons (panel (a)) and fermions (panel (b)). We can see here clearly that for bosons, accretion time dominates over formation time for $\sigma_{\rm n\chi}>10^{-50}$ cm$^2$. In the case of fermions, the formation time dominates below $m_\chi\sim10^9$ GeV for higher $\sigma_{\rm n\chi}$ values. This explains the general trends observed in Figs.~\ref{NSboson} and \ref{NSfermion}.

\begin{figure}[h!]
    \centering
\subfigure[Bosonic DM]{\includegraphics[width=2.5in]{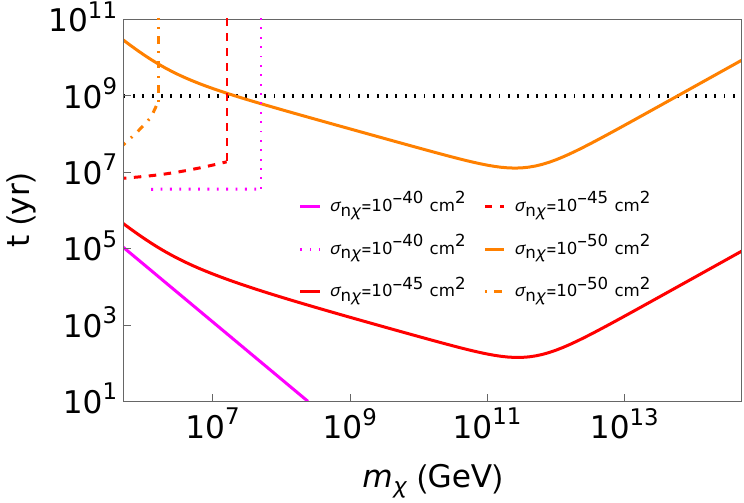}}
\hspace{0.01\textwidth}
\subfigure[Fermionic DM]{\includegraphics[width=2.5in]{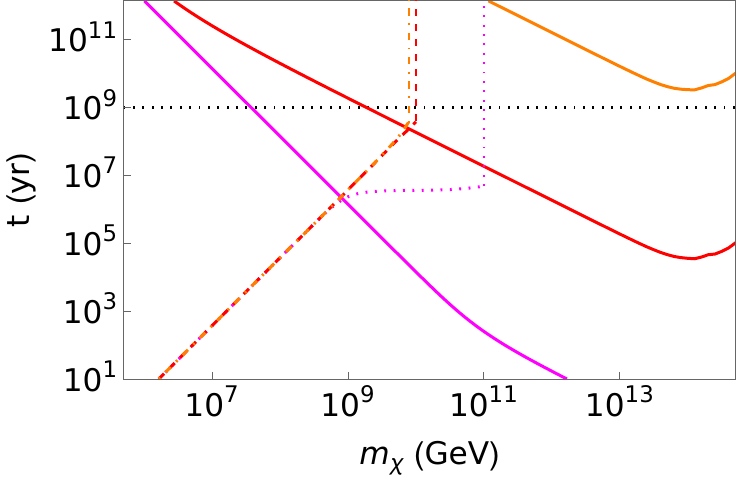}}
    
\caption {The formation timescales $t_0$ (solid curves) and the accretion timescales $t_{\rm acc}$ (dashed curves) plotted separately for $\rho_{\rm disk}=0.4$ GeV.cm$^{-3}$ and different $\sigma_{\rm n\chi}$. The accretion timescale is important at higher $\sigma_{\rm n\chi}$ values because it exceeds the formation timescale. Furthermore, the divergence of accretion timescale at certain $m_\chi$ also terminate the total timescale as in Figs.~\ref{NSboson} and \ref{NSfermion}.}
    \label{timescale-split-NS}
\end{figure}

\begin{figure}[h!]
    \centering
\subfigure[For Galactic disk ($\rho_{\rm disk}=0.4$ GeV.cm$^{-3}$)]{\includegraphics[width=2.1in]{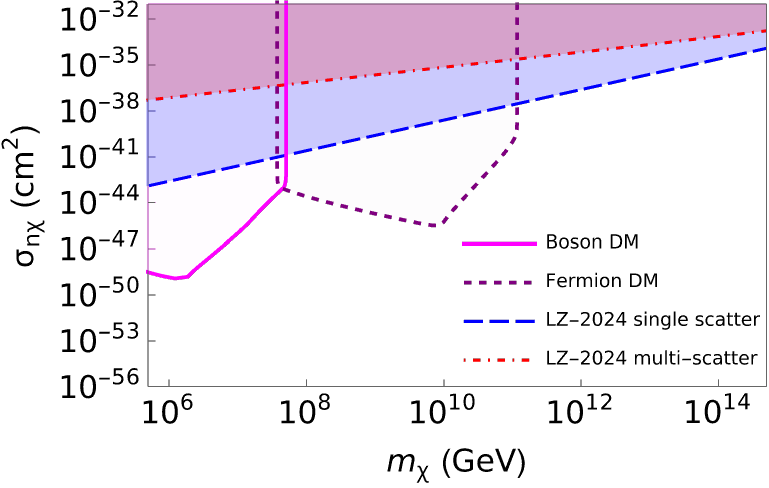}}
\hspace{0.01\textwidth}
\subfigure[For Galactic bulge ($\rho_{\rm bulge}=7\times10^4$ GeV.cm$^{-3}$)]{\includegraphics[width=2.1in]{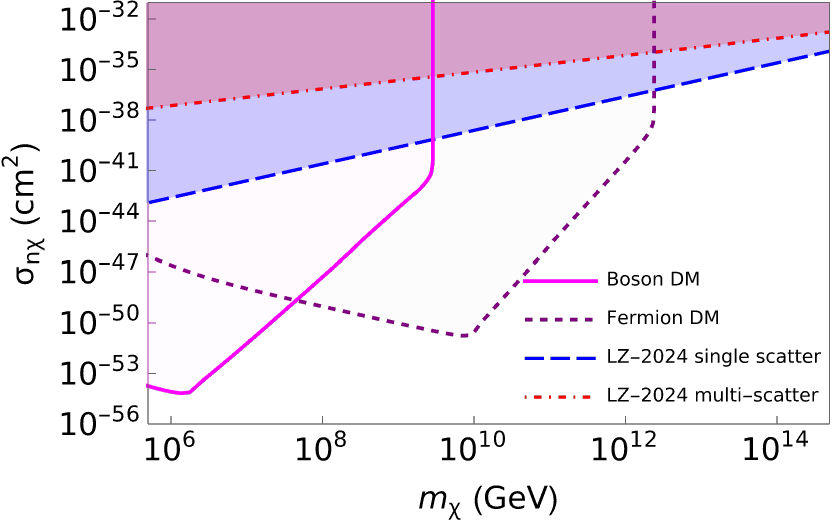}}
\hspace{0.01\textwidth}
\subfigure[For globular cluster ($\rho_{\rm GC}=10^3$ GeV.cm$^{-3}$)]{\includegraphics[width=2.1in]{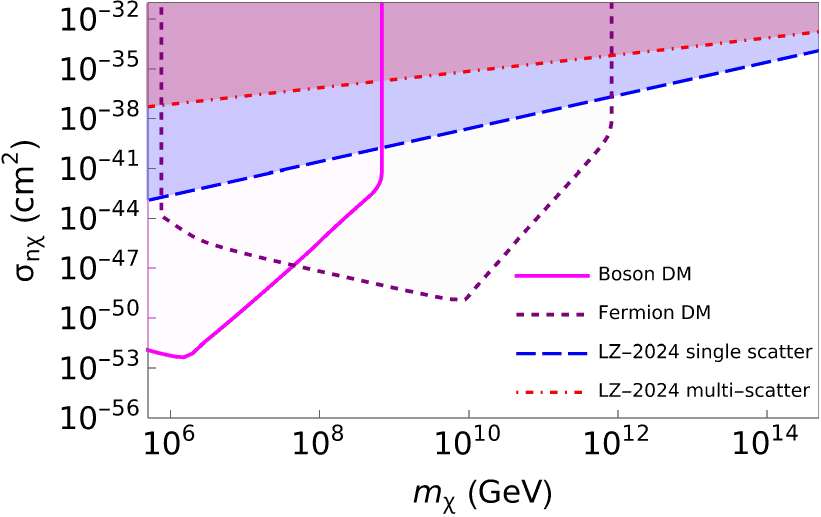}}

\caption{Constraints on the DM-nucleon scattering cross section $\sigma_{\rm n\chi}$, as a function of the DM particle mass $m_\chi$, obtained by setting transmutation time equal to the average age $T_{\rm MSP}$. The region above the curves corresponding to the bosonic and fermionic DM-imposed constraints (solid magenta and purple dashed curves) is the region where $t_{\rm d}<T_{\rm MSP}$. This is the region where DM interaction for the corresponding $m_\chi$ should have transmuted most MSPs of age $\ge1$ Gyr. Below the curve are the possible $\sigma_{\rm n\chi}$ values for which transmutation has yet to occur. The shaded regions represent the regions excluded in the $m_\chi-\sigma_{\rm n\chi}$ space by the LZ experimental data~\cite{LZ-2024}. 
The constraints are derived using Eq.~\eqref{td} For details, refer to section \ref{NS}.}
    \label{m-sigma}
\end{figure}

Assuming that a typical MSP has an age of $1$ Gyr, we can constrain $\sigma_{\rm n\chi}$ for DM mass in the range $10^6 \,{\rm GeV}<m_\chi<10^{14}\,{\rm GeV}$. 
For three different regions with DM density $\r_{\rm disk}\sim0.4\,{\rm GeV. cm}^{-3}$, $\r_{\rm bulge}\sim7\times10^4\,{\rm GeV. cm}^{-3}$ and $\r_{\rm GC}\sim10^3\,{\rm GeV. cm}^{-3}$, panels (a), (b), and (c) of Fig.~\ref{m-sigma} shows the $m_\chi-\sigma_{\rm n\chi}$ space with the constraint on $\sigma_{\rm n\chi}$ plotted for bosonic and fermionic DM particles using Eq.~\eqref{td}. 
For comparison, the LZ (LUX-ZEPLIN) experimental constraints~\cite{LZ-2024} on the DM-nucleon interaction cross-section are also shown in the shaded form. The blue dashed line enclosing the blue shaded region represents the extrapolation of the WIMP search results to higher DM masses~\cite{LZ-2024}, whereas the red dot-dashed line enclosing the red shaded region is the actual experimental constraint curve for multiply interacting massive particles (MIMPs)~\cite{LZ-2024, Bramante_MIMP_2018}.
In the case of MSPs, we can directly compare the results of Model 1 of~\cite{LZ-2024}, where, in spin-independent DM-nuclear scattering, the total DM-nucleus cross-section is coherently enhanced with respect to the DM-nucleon cross-section. 
For $\r_{\rm disk}$, fermionic DM provides no $\sigma_{\rm n\chi}$ constraint below $m_\chi\sim10^7$ GeV, and bosonic DM constraints are not applicable above $m_\chi\sim10^8$ GeV.  
Similarly for $\rho_{\rm GC}$ (panel (c)), fermionic DM does not provide any constraints below $\sim10^6$ GeV. 
One can look at the constraint curves as two separate segments. The left segment that runs diagonally from the left-top side to the right-bottom side is determined primarily by the decreasing formation time ($t_0$). The second segment, which extends from the endpoint of the first segment and runs through $m_\chi$ values to the right-top corner almost diagonally, is determined by the growth time $t_{\rm acc}$. 
The left cutoff seen in the case of fermions is due to the formation time exceeding $T_{\rm MSP}$ and the cutoff on the right edge is determined by $\dot{m}=0$ condition, as shown in Fig.~\ref{MdotNS}.
The constraints obtained from these calculations provide a significant improvement over the regions already excluded by the LZ experiment.

\begin{figure}
    \centering
    \includegraphics[width=2.5in]{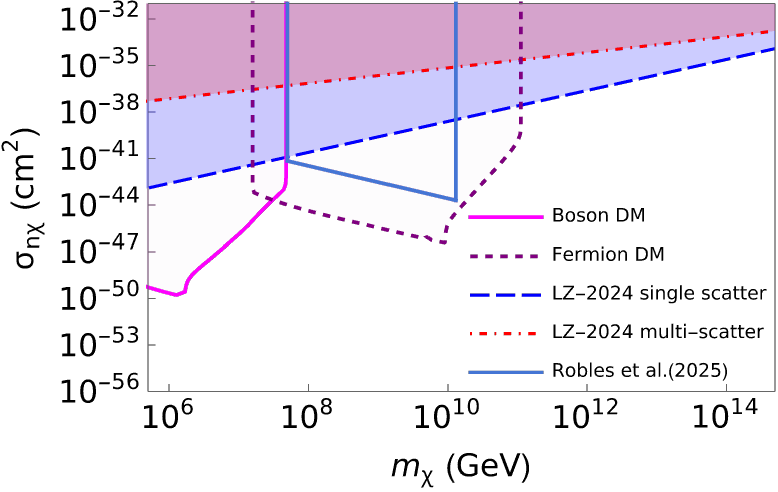}
    \caption{Comparison of the constraints obtained using Eq.~\eqref{td} with those reported in Ref.~\cite{Robles_2025} for the MSP PSR J0437$-$4715 of mass $1.4\,M_\odot$ and age $6.7$ Gyr. The dashed purple and solid blue curves represent the findings of the current work and those of Ref.~\cite{Robles_2025}, respectively, concerning fermionic dark matter (DM). In contrast, the solid magenta curve illustrates the constraint derived in the present work for bosonic DM. See Sec.~\ref{NS} for details.}
    \label{NS-comparison}
\end{figure}

 Fig.~\ref{NS-comparison} compares the constraints obtained in the $m_\chi-\sigma_{\rm n\chi}$ plane using Eq.~\eqref{td} for the MSP PSR J0437$-$4715 (with mass $\sim1.4,M_\odot$, age $6.7$ Gyr, and located in the Galactic disk) with those reported in \cite{Robles_2025}. For fermionic DM, the dashed purple and blue solid curves correspond to the present work and Ref.~\cite{Robles_2025}, respectively.
The difference in the order of magnitude of the obtained constraints arises from the difference in the formation and accretion timescale calculations performed in~\cite{Robles_2025} which include relativistic corrections. 
Another noticeable difference is the behavior of the exclusion boundary at large $m_\chi$. In ~\cite{Robles_2025}, the constraint terminates abruptly once Hawking evaporation dominates the EBH evolution. In contrast, our constraints extend over a limited region of the $(m_\chi,\sigma_{\rm n\chi})$ parameter space, where the EBH growth is still sustained by the continued DM accretion term. As discussed in Sec.~\ref{NS} and illustrated in panel (a) of Fig.~\ref{MdotNS} (see, also panel (a) of Fig.~\ref{NSfermion}), the DM accretion term remains non-negligible only within this limited region, leading to a gradual cutoff of the exclusion boundary. In addition, our formalism also enables us to derive the corresponding constraints for bosonic DM, shown by the solid magenta curve in Fig.~\ref{NS-comparison}.

\subsection{\label{WD} White dwarfs}

WDs typically have a longer lifespan than MSPs, and the oldest of them ($T_{\rm WD}>10$ Gyr) are found in globular clusters~\cite{Torres_WD_glob_cluster}, indirect evidence for $\sim10$ Gyr old WDs in the Galactic bulge~\cite{Calamida_2014_GB_WD}, as well as the Galactic disk~\cite{Fantin_2019_disk_WD}.
Substituting a typical $\rho \approx 10^{9}$ kg.m$^{-3}$ and core temperature $\T\sim10^7$ K~\cite{acevedo_supernovae_2019} in Eq.~\eqref{self}, we obtain,
\begin{eqnarray}
 N_{\rm self}^{\rm WD}=1.8 \times 10^{49} \left(\frac{100~{\rm GeV}}{m_{\chi}}\right)^{5/2}\left(\frac{\T}{10^8~{\rm K}}\right)^{3/2}.
 \label{nselfWD}
\end{eqnarray}
Now, we fix some WD parameters as follows. We consider WDs with an average mass of $0.6 M_\odot$ and radius $6000$ km (corresponding to $\sigma_{\rm crit}^{\rm WD}\approx10^{-39}{\rm cm}^2$) in the Galactic disk ($\rho_{\rm disk}=0.4$ GeV.cm$^{-3}$), Galactic bulge ($\rho_{\rm bulge}=7\times 10^4$ GeV.cm$^{-3}$), and globular cluster ($\rho_{\rm GC}=10^3$ GeV.cm$^{-3}$) regions.

\begin{figure}[h!]
    \centering
\subfigure[For Galactic disk ($\rho_{\rm disk}=0.4$ GeV.cm$^{-3}$)]{\includegraphics[width=2.5in]{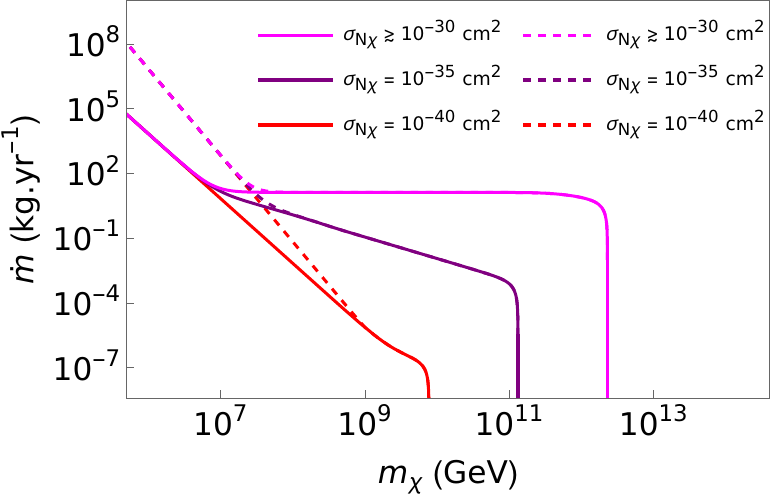}}
\hspace{0.01\textwidth}
\subfigure[For Galactic bulge ($\rho_{\rm bulge}=7\times10^4$ GeV.cm$^{-3}$)]{\includegraphics[width=2.5in]{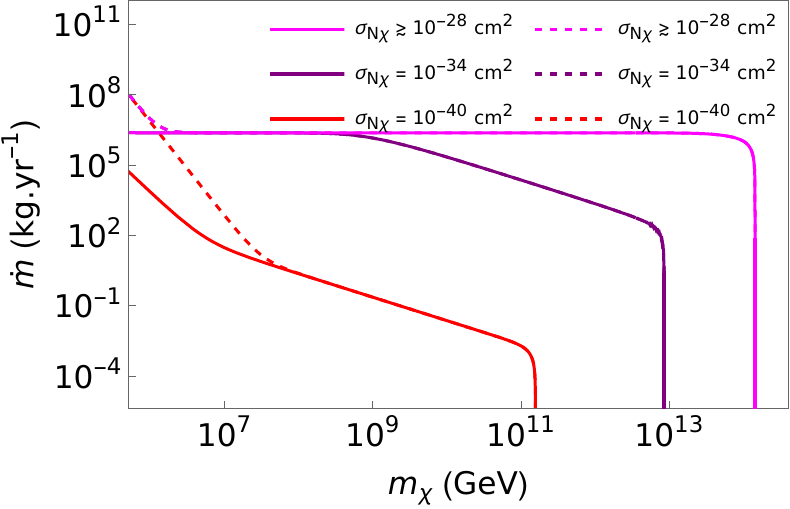}}
\hspace{0.01\textwidth}
\subfigure[For Globular cluster ($\rho_{\rm GC}=10^3$ GeV.cm$^{-3}$)]{\includegraphics[width=2.5in]{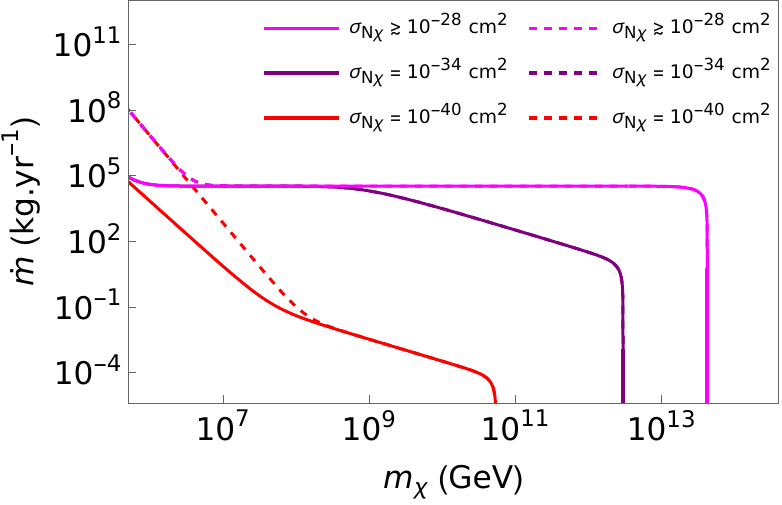}}
\hspace{0.01\textwidth}
\subfigure[$\dot{m}$ considered to be independent of $\rho_{\rm DM}$, i.e., $k_2 \rightarrow 0$]{\includegraphics[width=2.5in]{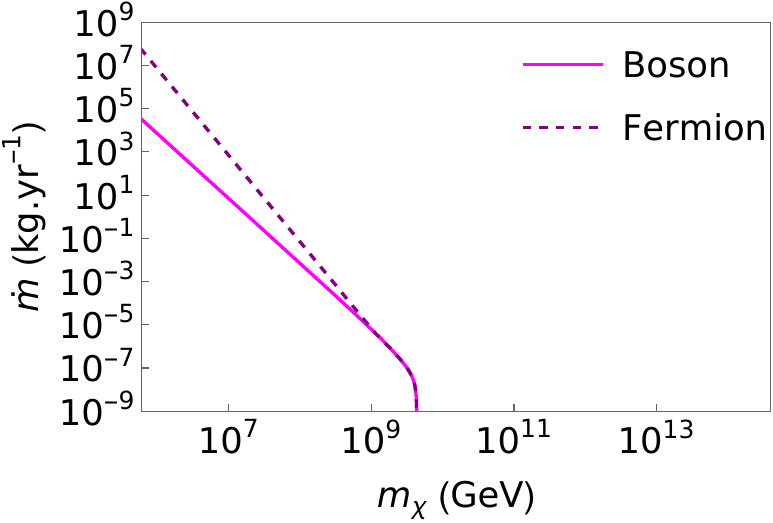}}

\caption{Similar to Fig.~\ref{MdotNS}, but showing the initial growth rate of an EBH located inside a WD of mass $0.6 M_\odot$ and radius $6000$ km. For details refer to Sec.~\ref{WD}.}
    \label{MdotWD}
\end{figure}

Similarly to MSPs, we consider the initial growth rate of the EBH formed inside a WD, as a function of $m_\chi$, to mark the domain of Eq.~\eqref{td}. Here, we see a trend similar to that of an MSP (panels (a), (b) and (c) of Fig.~\ref{MdotWD}) wherein, for a higher $\sigma_{\rm N\chi}$ ($\sigma_{\rm N\chi}\gg\sigma_{\rm crit}^{\rm WD}$), we see a large range of $m_\chi$ for which $\dot{m}$ is dominated by DM accretion but as $\sigma_{\rm N\chi}$ decreases, the DM accretion rate tends to zero and $\dot{m}$ becomes governed entirely by baryon accretion and Hawking radiation. Furthermore, above a certain $m_\chi$, EBH formed from both fermionic and bosonic DM particles have the same $\dot{m}$. This is simply because the initial mass of EBH above $m_{\chi} \sim10^8$ GeV is determined by the collapse criterion $N>N_{\rm self}^{\rm WD}$ for both fermionic and bosonic DM particles (see Fig.~\ref{fm0}).
A comparison can be made with a case where the continued capture of DM is not taken into account (panel (d) of Fig.~\ref{MdotWD}), highlighting the importance of considering all three contributing factors to the growth rate in Eq.~\eqref{tot1} and the resulting timescale formula, Eq.~\eqref{td}. Without the three terms involved in Eq.~\eqref{tot}, one would expect that the EBH formed from DM with $m_\chi\gtrsim10^{9}$ GeV ($M_{\rm crit1}\lesssim 10^{11}$ kg) would simply evaporate and not transmute the host. 
In stark contrast, with the inclusion of the DM accretion rate in the calculation of $\dot{m}$, we see that the EBH formed from DM particles of up to $\sim10^{12}$ GeV (i.e. EBH of mass $M_0=M_{\rm crit}\sim10^7$ kg) can still grow (rather slowly) and transmute its host located in the Galactic disk, for $\sigma_{\rm N\chi}\gg\sigma_{\rm crit}^{\rm WD}$ (panel (a) of Fig.~\ref{MdotWD}). For a WD located in the Galactic bulge, EBH formed from DM particles of mass up to $\sim10^{14}$ GeV (panel (b)), thus $M_0=M_{\rm crit}\approx4\times10^4$ kg can transmute it, and for a WD located in the GC, these values are $\sim10^{13}$ GeV (panel (c))  and $M_{\rm crit}\approx3\times10^5$ kg, respectively.

\begin{figure}[h!]
    \centering
    \includegraphics[width=2.5in]{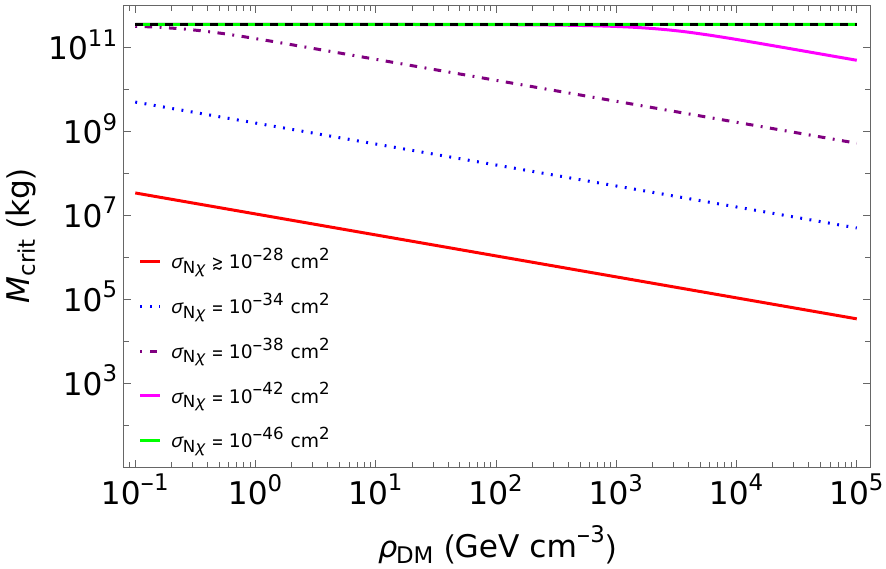}
    \caption{$M_{\rm crit}$ given by Eq.~\eqref{mcrit} inside a WD as a function of the DM density $\rho_{\rm DM}$ in its surroundings and $\sigma_{\rm N\chi}$. The black dashed line indicates the $M_{\rm crit1}$ given by Eq.~\eqref{mcr}. For details refer to Sec.~\ref{WD}.}
    \label{Mcrit-WD}
\end{figure}

Now we can also look at the minimum mass of an EBH that it needs to have to transmute its host, given by Eq. ~\eqref{mcrit}, inside a WD as a function of the DM density $\rho_{\rm DM}$ in its surroundings. Fig.~\ref{Mcrit-WD} shows, as in the case of MSPs, in WDs, there is a trend of increasing $M_{\rm crit}$ with decreasing $\r_{\rm DM}$. In the case of WDs located in the Galactic disk, for example, the DM particles (bosons or fermions) must have $m_\chi\lesssim10^{12}$ GeV to transmute the host into a BH.
For lower $\sigma_{\rm N\chi}$ one can see that the $M_{\rm crit}$ curves approach $M_{\rm crit1}$ for lower $\rho_{\rm DM}$ and eventually for $\sigma_{\rm N\chi}<10^{-45}$ cm$^2$, the curves coincide since $k_2\rightarrow0$.

\begin{figure}[h!]
 \centering
\subfigure[For Galactic disk ($\rho_{\rm disk}=0.4$ GeV.cm$^{-3}$)]{\includegraphics[width=2.1in]{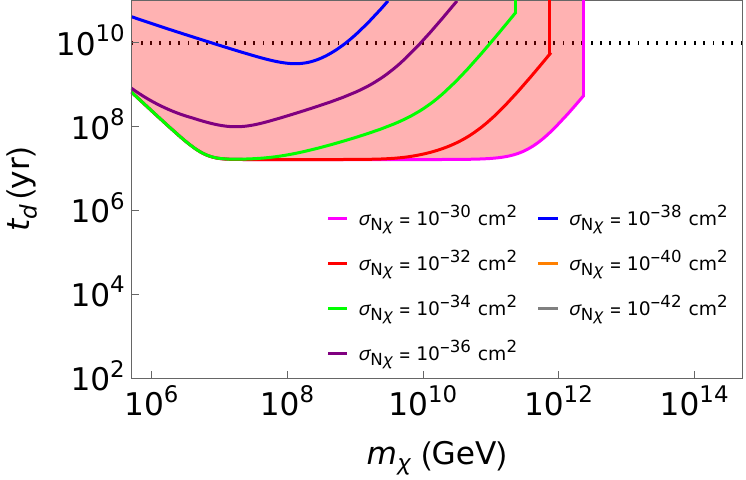}}
\hspace{0.01\textwidth}
\subfigure[For Galactic bulge ($\rho_{\rm bulge}=7 \times 10^4$ GeV.cm$^{-3}$)]{\includegraphics[width=2.1in]{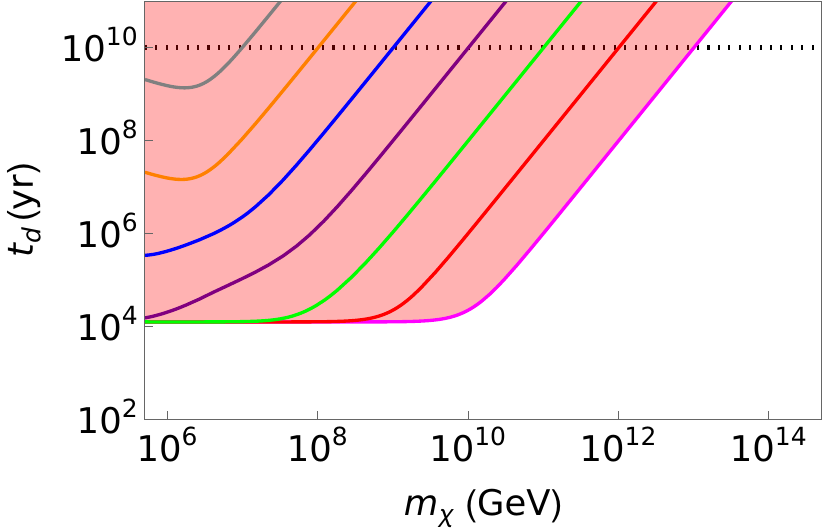}}
\hspace{0.01\textwidth}
\subfigure[For globular cluster ($\rho_{\rm GC}=10^3$ GeV.cm$^{-3}$)]{\includegraphics[width=2.1in]{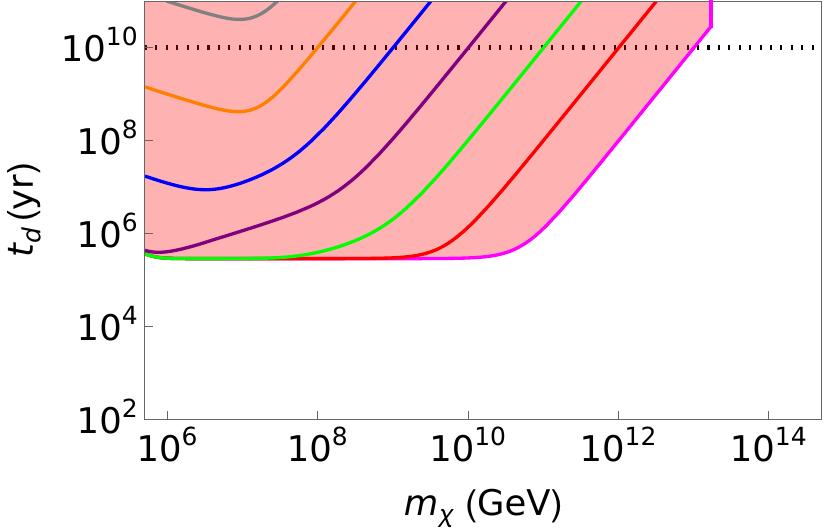}}
\hspace{0.01\textwidth}
\subfigure[For Galactic disk ($\rho_{\rm disk}=0.4$ GeV.cm$^{-3}$)]{\includegraphics[width=2.1in]{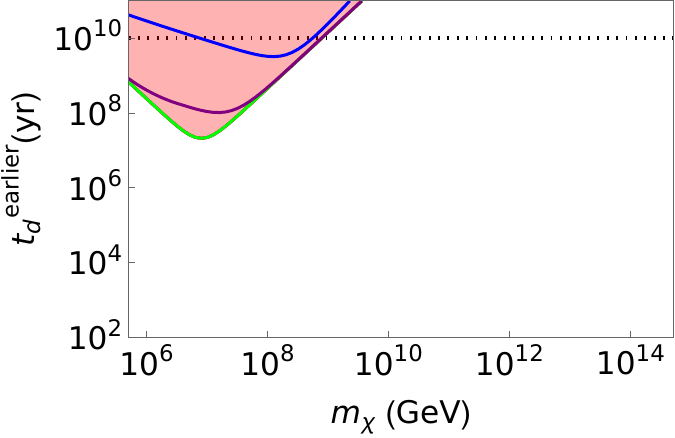}}
\hspace{0.01\textwidth}
\subfigure[For Galactic bulge ($\rho_{\rm bulge}=7 \times 10^4$ GeV.cm$^{-3}$)]{\includegraphics[width=2.1in]{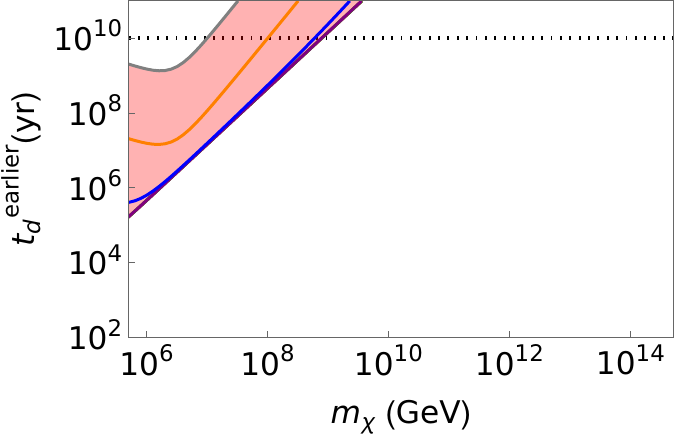}}
\hspace{0.01\textwidth}
\subfigure[For globular cluster ($\rho_{\rm GC}=10^3$ GeV.cm$^{-3}$)]{\includegraphics[width=2.1in]{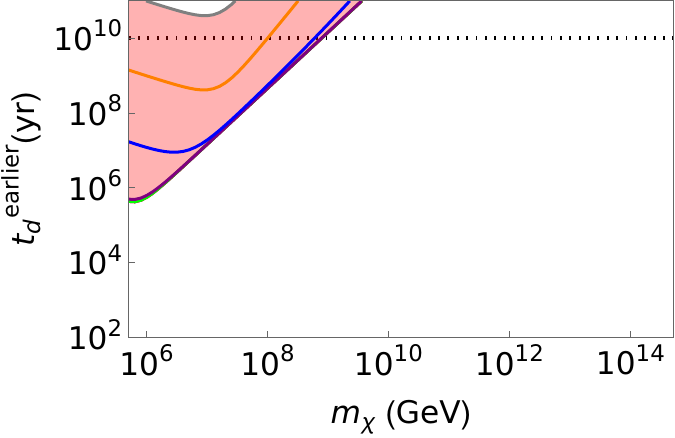}}

\caption{\label{WDboson} Transmutation time $t_{\rm d}$ as a function of the DM particle mass $m_\chi$ for a WD located in different galactic environments. The shaded regions correspond to DM–nucleon scattering cross sections in the range $\sigma_{\rm N\chi} \in [10^{-42},10^{-30}] {\rm cm}^2$. Panels (a)–(c) are obtained using Eq.~\eqref{td}, whereas panels (d)–(f) are computed using the earlier prescription given by Eq.~\eqref{earlier}. We assume that the WD captures asymmetric DM particles with a velocity dispersion $\bar v = 220~{\rm km~s^{-1}}$, and that the sound speed inside the WD is $c_s = 0.016c$~\cite{balberg_properties_2000}. The black dotted line denotes the typical age of WDs, $T_{\rm WD} \sim 10~{\rm Gyr}$. The apparent cutoff at certain values of $m_\chi$ in panels (a), (b), and (c) arises due to the divergence $t_{\rm d} \rightarrow \infty$ when the initial mass satisfies $M_0 = M_{\rm crit}$, corresponding to $\dot m \rightarrow 0$. Significant differences are evident between the transmutation times $t_{\rm d}$ and $t_{\rm d}^{\rm earlier}$. For further discussion, see Sec.~\ref{WD}.}

\end{figure}

\begin{figure}[h!]
 \centering
\subfigure[For Galactic disk ($\rho_{\rm disk}=0.4$ GeV.cm$^{-3}$)]{\includegraphics[width=2.1in]{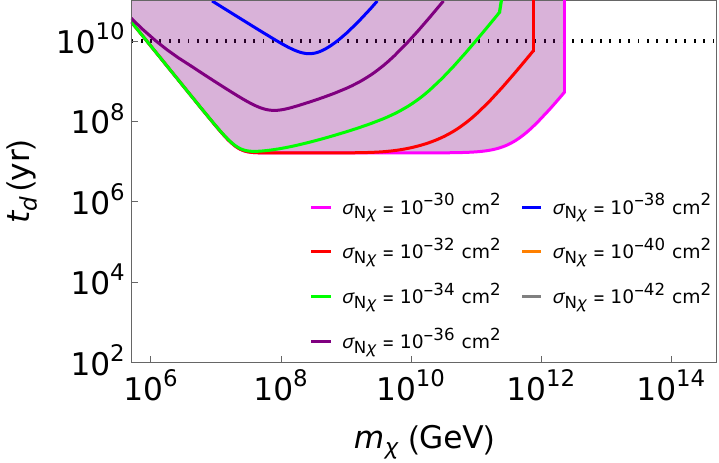}}
\hspace{0.01\textwidth}
\subfigure[For Galactic bulge ($\rho_{\rm bulge}=7 \times 10^4$ GeV.cm$^{-3}$)]{\includegraphics[width=2.1in]{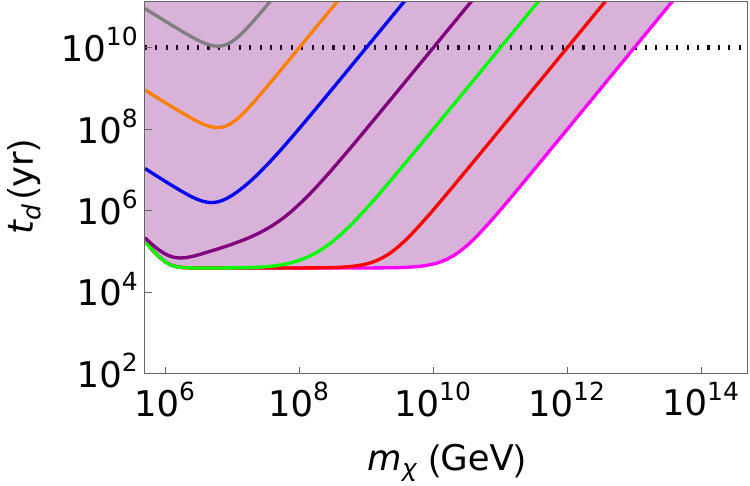}}
\hspace{0.01\textwidth}
\subfigure[For globular cluster ($\rho_{\rm GC}=10^3$ GeV.cm$^{-3}$)]{\includegraphics[width=2.1in]{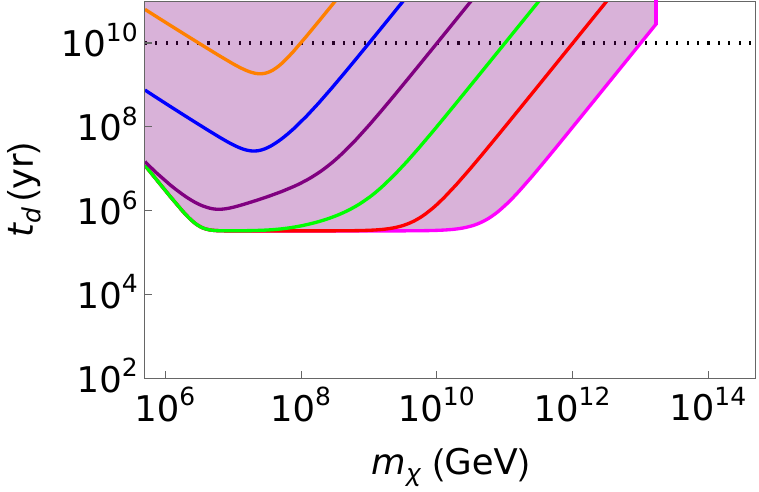}}
\hspace{0.01\textwidth}
\subfigure[For Galactic disk ($\rho_{\rm disk}=0.4$ GeV.cm$^{-3}$)]{\includegraphics[width=2.1in]{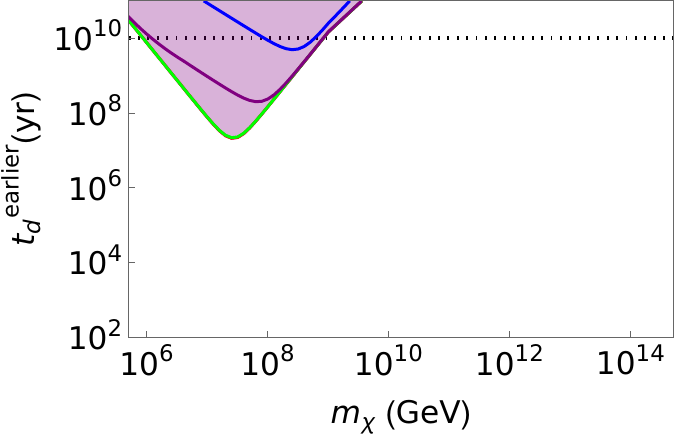}}
\hspace{0.01\textwidth}
\subfigure[For Galactic bulge ($\rho_{\rm bulge}=7 \times 10^4$ GeV.cm$^{-3}$)]{\includegraphics[width=2.1in]{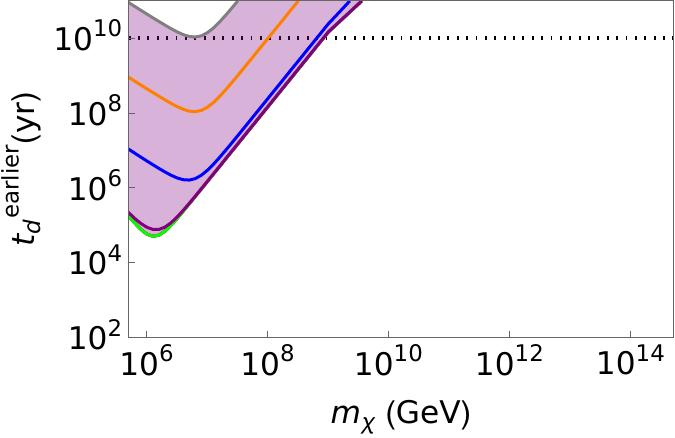}}
\hspace{0.01\textwidth}
\subfigure[For globular cluster ($\rho_{\rm GC}=10^3$ GeV.cm$^{-3}$)]{\includegraphics[width=2.1in]{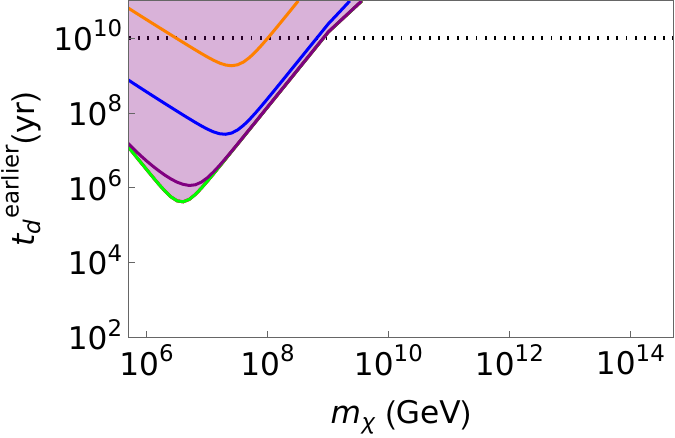}}

\caption{\label{WDfermion} Same as Fig.~\ref{WDboson} but for asymmetric fermionic DM. Notice that above $m_\chi\sim 10^8$ GeV, the $t_{\rm d}$, as well as the $t_{\rm d}^{\rm earlier}$ curves for fermionic DM, start resembling their counterparts in bosonic DM. The resemblance is explained by the initial mass $M_0$ being independent of the nature of DM above $m_\chi\sim 10^8$ GeV, as can be seen in Panel (b) of Fig.~\ref{fm0}. For details refer to Sec.~\ref{WD}.}

\end{figure}

\begin{figure}[h!]
    \centering
\subfigure[Bosonic DM]{\includegraphics[width=2.5in]{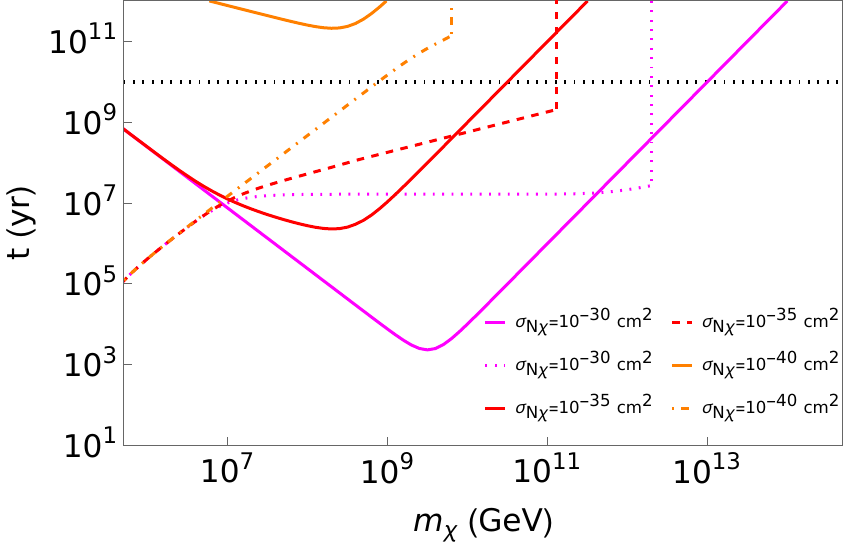}}
\hspace{0.01\textwidth}
\subfigure[Fermionic DM]{\includegraphics[width=2.5in]{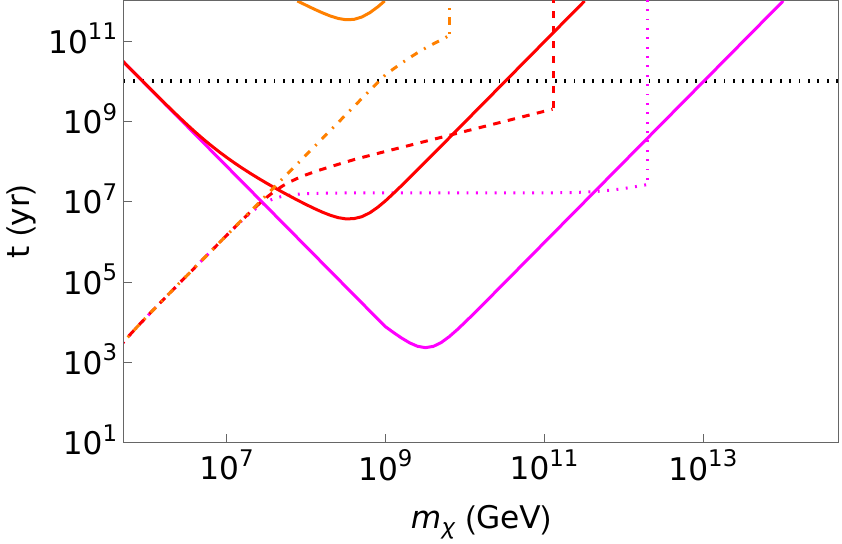}}
    
\caption{The formation timescales $t_0$ (solid curves) and the accretion timescales $t_{\rm acc}$ (dashed curves) plotted separately for $\rho_{\rm disk}=0.4$ GeV.cm$^{-3}$ and different $\sigma_{\rm N\chi}$. The accretion timescale is observed to be important at higher $\sigma_{\rm N\chi}$ as it exceeds the formation timescale in a certain range of $m_\chi$. Furthermore, the divergence of the accretion timescale at certain $m_\chi$ also terminates the total timescale, as shown in Fig.~\ref{WDboson} and \ref{WDfermion}.} 
    \label{timescale-split-WD}
\end{figure}

Now we can apply Eq.~\eqref{td} for different $\r_{\rm DM}$ and $\sigma_{\rm N\chi}\in[10^{-42},10^{-30}] {\rm cm}^2$.
We find that for both bosonic and fermionic DM particles, Eq.~\eqref{td} gives drastically different results compared to Eq.~\eqref{earlier}. 
In Fig.~\ref{WDboson} and Fig.~\ref{WDfermion} panels (a), (b), and (c) show the transmutation timescales $t_{\rm d}$ plotted for the above-mentioned parameter values for bosonic and fermionic DM particles, respectively, using Eq.~\eqref{td}. Panels (d), (e), and (f) show the timescales calculated and plotted for the same parameter values, using Eq.~\eqref{earlier}.
A noticeable feature of these figures is that, for higher DM densities, such as those in the Galactic bulge and globular clusters, the curves corresponding to different values of $\sigma_{\rm N\chi}$ intersect the black dotted line at nearly the same value of $m_\chi$. Consequently, the resulting $m_\chi$--$\sigma_{\rm n\chi}$ exclusion curves are very similar over the range $m_\chi\in[10^8,10^{13}]~\mathrm{GeV}$, despite the order-of-magnitude differences in the ambient DM density.
As in the case of MSPs, there exists a range of $m_\chi$ and $\sigma_{\rm N\chi}$ for which the transmutation timescale $t_{\rm d}$ is nearly independent of the initial EBH mass $M_0$ (or equivalently $m_\chi$). However, unlike MSPs, this behavior is present for both bosonic and fermionic DM. This is because, above $m_\chi\sim10^8$ GeV, the initial EBH mass is determined by the self-gravitation criterion (Eq.~\eqref{self}) for both bosonic and fermionic DM, as illustrated in panel (b) of Fig.~\ref{fm0}.
Fig.~\ref{timescale-split-WD} shows the formation and accretion timescales separately for $\rho_{\rm disk}=0.4$ GeV.cm$^{-3}$ and different values of $\sigma_{\rm N\chi}$, for bosonic (panel (a)) and fermionic (panel (b)) DM. It is seen that the formation timescale in WDs decreases at lower $m_\chi$ and subsequently increases with increasing $m_\chi$. The figure clearly identifies the regions in which either the formation timescale ($t_0$) or the accretion timescale ($t_{\rm acc}$) dominates the total transmutation timescale. Consequently, the resulting constraint on $\sigma_{\rm n\chi}$ is determined by whichever contribution dominates when the total transmutation timescale becomes comparable to the characteristic WD age, $T_{\rm WD}\sim10$ Gyr.

\begin{figure}[h!]
    \centering
\subfigure[For Galactic disk ($\rho_{\rm disk}=0.4$ GeV.cm$^{-3}$)]{\includegraphics[width=2.1in]{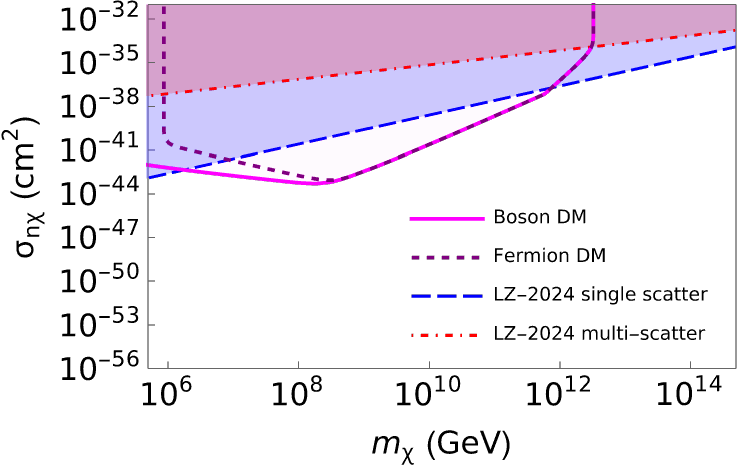}}
\hspace{0.01\textwidth}
\subfigure[For Galactic bulge ($\rho_{\rm bulge}=7\times10^4$ GeV.cm$^{-3}$)]{\includegraphics[width=2.1in]{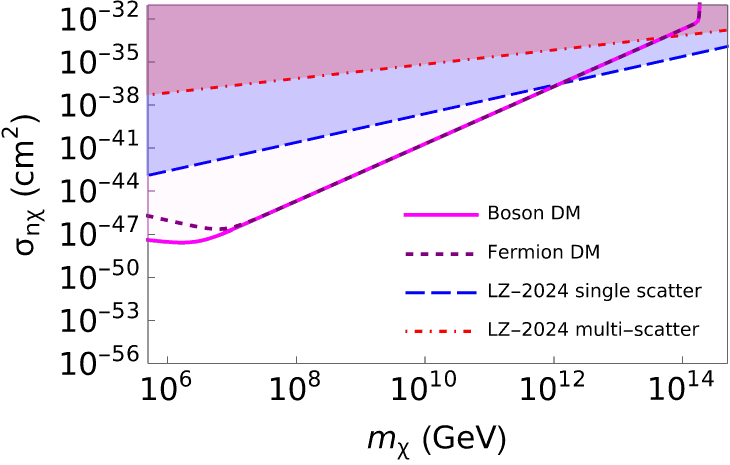}}
\hspace{0.01\textwidth}
\subfigure[For globular cluster ($\rho_{\rm GC}=10^3$ GeV.cm$^{-3}$)]{\includegraphics[width=2.1in]{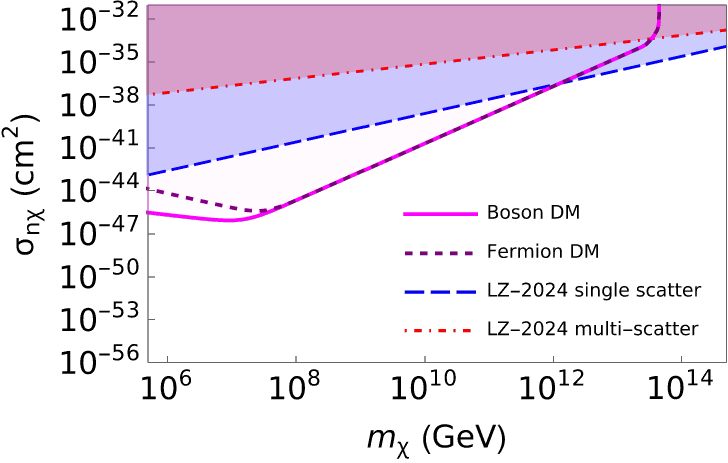}}
    
\caption{Similar to Fig.~\ref{m-sigma} but for WDs with age $T_{\rm WD}\sim10$ Gyr. The region above the curves corresponding to the bosonic and fermionic DM-imposed constraints is the region where one expects $t_{\rm d}<T_{\rm WD}$. This is the region where the DM interaction for the corresponding $m_\chi$ should have transmuted most WDs of age $\ge10$ Gyr. Below the curve are the possible $\sigma_{\rm n\chi}$ values for which transmutation has yet to occur. The shaded regions represent the regions excluded in the $m_\chi-\sigma_{\rm n\chi}$ space by the LZ experimental data~\cite{LZ-2024}. 
For discussion refer to Sec.~\ref{WD}.}
    \label{m-sigma-wd}
\end{figure}

Fig.~\ref{m-sigma-wd} shows the constraints placed on $\sigma_{\rm n\chi}$ for $m_\chi\in[5\times10^5, 5\times10^{14}]\,{\rm GeV}$, considering the average age of WDs to be $T_{\rm WD}\sim10$ Gyr.~\footnote{While there are no direct measurements of faint WDs which suggest an age $\sim10$ Gyr in the galactic bulge, analyses~\cite{Calamida_2014_GB_WD, NASA} have confirmed the presence of a WD cooling sequence in the Galactic bulge and its consistency with an old ($\sim10$ Gyr) stellar population. Therefore the constraints imposed by our calculations should be interpreted in light of this.}
To compare with the constraints placed by the LZ experiment (which gives constraints on the DM-nucleon cross-section), the obtained cross-section (for a carbon-oxygen WD, the nuclei have an average mass number of $\sim14$ u.) from the calculation has been scaled according to the A$^4$ scaling relation of model 1 discussed in~\cite{LZ-2024}. 
We notice that the constraints obtained from WDs on $\sigma_{\rm n\chi}$ are qualitatively different from those placed by MSPs. Furthermore, we notice that the constraints placed by bosonic and fermionic DM for higher DM densities (in the Galactic bulge and globular clusters) are more stringent at lower $m_\chi$ than those obtained from lower DM densities (in the Galactic disk). 
Another interesting feature that appears in the constraint curve in panel (a) is the change in slope of the diagonal curve from $m_\chi\sim10^{11}$ GeV to $m_\chi\sim10^{12}$ GeV, which is not found in panels (b) and (c). The accretion timescale (or, in turn, the total timescale) diverges at different $m_\chi$ values for different $\sigma_{\rm N\chi}$ values, which, in the case of the Galactic disk, occurs at the order of $T_{\rm WD}=10$ Gyr between $m_\chi\sim10^{11}$ GeV and $m_\chi\sim10^{12}$ GeV (refer to Panel (a) in Figs.~\ref{WDboson} and \ref{WDfermion}). 
Therefore, in this range of $m_\chi$ the timescale curves corresponding to different $\sigma_{\rm N\chi}$ cross the $T_{\rm WD}=10$ Gyr limit more closely to each other than in the case of the Galactic bulge and globular cluster, and hence the constraint curve becomes steeper here.
While a similar feature is barely visible in panel (c), it is not seen in panel (b) since in higher $\rho_{\rm DM}$ regions, timescale diverges at higher orders due to the additional DM accretion rate.
Because we have used the condition $t_{\rm d}>T_{\rm WD}$ where $T_{\rm WD}\sim10$ Gyr, it is natural to consider possible changes due to the host's temperature drop during its lifetime. This mainly affects the DM accretion rate in Eqs. \eqref{tot1} and \eqref{tot2} (see Eq. (4) and (22) in~\cite{steigerwald_revisiting_2022}). However, we can check the effect to be small enough that it does not significantly change the timescale. Therefore, the constraints shown in panels (a), (b), and (c) in Fig.~\ref{m-sigma-wd} remains valid.

\begin{figure}
    \centering
    \includegraphics[width=2.5in]{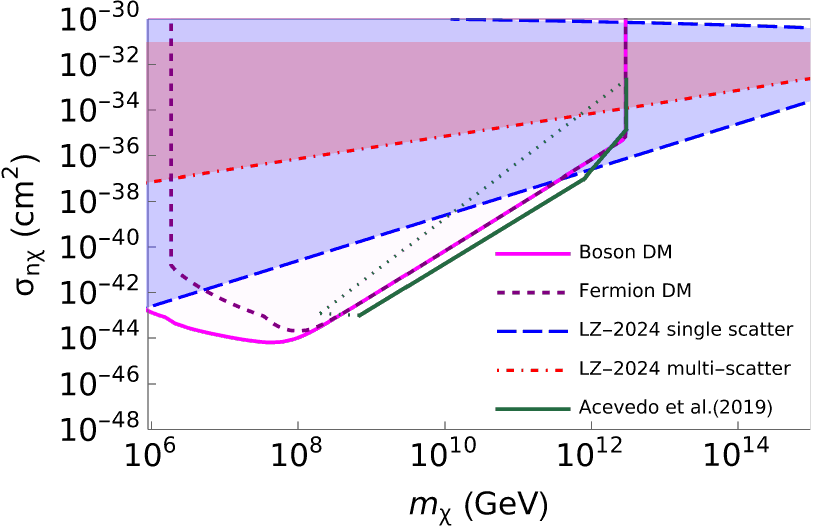}

    \caption{Comparison of the constraints obtained using Eq.~\eqref{td} with those reported in Ref.~\cite{acevedo_supernovae_2019} for the WD SDSS J160420.40$+$055542.3 of mass $\sim1.4\,M_\odot$, radius $2.5\times10^3$ km, and age $3$ Gyr. The dashed purple and solid green curves correspond to the fermionic DM constraints obtained in the present work and in Ref.~\cite{acevedo_supernovae_2019}, respectively. The solid magenta curve shows the corresponding constraint obtained in the present work for bosonic DM. The dotted green curve denotes the WD ignition limit reported in Ref.~\cite{acevedo_supernovae_2019}, which is shown for completeness but is beyond the scope of the present work.  See Sec.~\ref{WD} for details.}
    \label{WD-comparison}
\end{figure}

Fig.~\ref{WD-comparison} shows a comparison between the constraints obtained in $m_\chi-\sigma_{\rm n\chi}$ space using Eq.~\eqref{td} for one old WD SDSS J160420.40 + 055542.3 with a mass $1.4\, M_\odot$ and an age of $3$ Gyr, and the constraints obtained in earlier works~\cite{acevedo_supernovae_2019} (see their Fig. 2). 
Although both studies derive constraints from the observed survival of an old WD, the theoretical criteria used to relate this observational constraint to the DM parameter space differ. In~\cite{acevedo_supernovae_2019}, the lower dotted segment of the green curve was primarily determined by the requirement that the DM accumulates into a self-gravitating core and collapses within the observed WD lifetime. 
The middle solid segment of the green curve, which is almost parallel to the constraint curves (solid magenta and dashed purple) from our calculations, was obtained under the condition that the EBH formed within the age of the WD. Its magnitude is slightly different from our calculations because we add the growth timescale to the DM-collapse time.
The upper solid segment of the green curve, which connects to our constraint curve, was obtained from the subsequent evolution of the newly formed black hole, determined by the competition between Hawking evaporation and continued DM feeding, and the surviving BH was assumed to eventually destroy the host WD. In contrast, our analysis does not constrain the parameter space using self-gravity or black hole survival conditions alone. 
Instead, once the collapse criterion is satisfied, we explicitly follow the complete evolution of the EBH by solving the growth equation, including baryonic accretion, Hawking evaporation, and continued DM feeding, and we require the total transmutation timescale, including both the formation and growth phases, to exceed the observed age of the WD. 
Consequently, the lower part of our exclusion curves is governed by the complete transmutation timescale rather than the self-gravity condition, whereas the steep upper branch naturally arises from the divergence of the transmutation timescale as the initial BH mass approaches the critical mass for sustained growth ($M_0 \rightarrow M_{\rm crit}$). We can see that this neatly matches the earlier constraints. 
The dotted green curve in the upper part of the figure shows the minimum $\sigma_{\rm n\chi}$ required for WD ignition according to the criterion derived in~\cite{acevedo_supernovae_2019}. As WD ignition is beyond the scope of the present work, we do not consider this criterion.
These differences in the underlying theoretical treatment lead to the slight quantitative differences between our exclusion curves and those of~\cite{acevedo_supernovae_2019}, despite both analyses employing the observed existence of old WDs as the primary astrophysical constraint.

\section{\label{sec6}Conclusion and discussion}
Taking into account accretion, evaporation, and capture rates simultaneously, we derived an exact analytical expression (Eq.~\ref{td}) for the timescale over which an astronomical object is transmuted into a black hole due to an EBH formed in its core through the capture of DM particles.
Although MSPs rotate rapidly, this rotation does not significantly modify the Bondi accretion rate or the resulting transmutation timescale~\cite{kouvaris_growth_2014, AHA_stall_ang_2025}. We further assume that the astronomical objects under consideration are not extremely magnetized.
Under these assumptions, using Eq.~\eqref{td} and requiring the transmutation timescale to exceed the typical ages of MSPs ($\gtrsim 1~\mathrm{Gyr}$) and WDs ($\gtrsim 10~\mathrm{Gyr}$), we derived constraints on the DM particle mass $m_\chi$ and the DM--nucleon scattering cross-section $\sigma_{\rm n\chi}$.

An interesting outcome of our analysis is that EBHs with initial masses as small as $\sim 10^{4}~\mathrm{kg}$ could undergo sustained growth and ultimately transmute their host objects. This behavior contrasts sharply with the expectations based solely on Bondi accretion. For comparison, it is well known that primordial black holes (PBHs) with masses $M_{\rm PBH}\lesssim 10^{12}~\mathrm{kg}$ have evaporation timescales comparable to the age of the Universe~\cite{hawking_particle_1975}, implying that lighter PBHs would not survive to the present epoch (see, however ~\cite{C_Chakraborty_22_gravmag}).
In the case of EBHs, considering only the Bondi accretion of host matter and Hawking radiation, one finds that EBHs with masses $\sim 10^{10}~\mathrm{kg}$ may still survive and grow within compact stars. However, when the possibility of continued DM feeding is included~\cite{acevedo_supernovae_2019, janish_type_2019}, EBHs with initial masses well below $M_0 \sim 10^{12}~\mathrm{kg}$ can remain stable and grow inside MSPs and WDs, eventually transmuting their hosts into black holes.
This opens the possibility of black holes with masses as low as $\sim 10^{4}~\mathrm{kg}$ existing in the Universe, albeit confined to the interiors of compact astrophysical objects such as NSs and WDs.

A few caveats are in order. Our calculations assume that the accretion of host matter onto the EBH can occur in the quantum regime (Unruh rate) or in a spherically symmetric Bondi flow, depending on the mass of the EBH. This approximation may not be fully adequate for rapidly rotating and/or highly magnetized WDs, where rotation and magnetic fields can suppress the accretion rate and thereby slow the growth of the EBH~\cite{AHA_stall_ang_2025,AHA_2025_MAT}. In such cases, the resulting transmutation timescales would be longer, allowing for values of the DM--nucleon scattering cross-section $\sigma_{\rm n\chi}$ that are modestly higher than those inferred from our fiducial assumptions. There can also be quantitative corrections arising from general relativistic effects, as well as from different prescriptions for the EBH formation and thermalization timescales, including the effects of nuclear reactions, alternative collapse criteria~\cite{Robles_2025}, and heat diffusion in WDs~\cite{steigerwald_revisiting_2022}.
Nevertheless, within the assumptions adopted in the present work regarding the EBH formation and growth processes, our results provide a physically transparent description of DM-induced transmutation in compact stars. The analytical framework developed here, together with its validation through numerical calculations, provides a useful benchmark for future studies that incorporate these additional physical effects in a more comprehensive treatment.

The transmutation timescale derived in this work can also be applied to low-mass, old stars, such as metal-poor main-sequence stars with masses $\lesssim 1M_\odot$. However, owing to their substantially lower central densities compared to MSPs and WDs, the capture rate of DM particles in these stars is highly suppressed. Consequently, the corresponding transmutation timescales typically exceed the age of the Universe for most values of $\sigma_{\rm n\chi}$, rendering these systems ineffective for placing competitive constraints on the DM–nucleon scattering cross-section.
Within the same framework, we can additionally obtain an analytical expression for the evaporation timescale of an EBH embedded in a generic astrophysical object, provided that the initial mass evolution satisfies $\dot m \leq 0$. The evaporation of an EBH may, in principle, induce observable effects on the host star. A detailed analysis of the evaporation timescale and its potential observational signatures is currently being prepared and will be presented elsewhere~\cite{Adarsha_2026}.

Finally, it has been argued that sufficiently strong repulsive DM self-interactions can inhibit EBH formation in the cores of MSPs~\cite{Dasgupta_2020}. In such scenarios, the black hole formation criterion can only be satisfied for sufficiently weak repulsive self-interactions. Consequently, the results presented here apply to ultra-heavy DM particles characterized by weak DM–baryon interactions and even weaker DM self-interactions.

\begin{acknowledgements}
HAA acknowledges Dr. TMA Pai Ph.D. scholarship program of Manipal Academy of Higher Education (MAHE). CC acknowledges the support of MAHE. We sincerely thank the referee for the constructive comments, which have significantly improved the quality of the paper.
\end{acknowledgements}

\onecolumngrid 

\appendix

\section{\label{ap1}Transmutation timescales in the special cases} 
\subsection{\label{ss1}Special case: $|\dot{m}_{\rm Hr}| \ll \dot{m}_{\rm acc}$ with $\dot{m}_{\rm DM} \ll \dot{m}_{\rm acc}$ }

If mass loss due to Hawking radiation is considered to be negligibly small, that is, $|\dot{m}_{\rm Hr}| \ll \dot{m}_{\rm acc}$, and not taken into account for the calculation of $t_{\rm d}$, one can equivalently set $k_3 \rightarrow 0$ (i.e., $k_4 \rightarrow k_2$) in Eq.~\eqref{td}. Thus, we obtain
\begin{eqnarray}
 t_{\rm d} \big|_{k_3 \rightarrow 0}  = t_0 + \frac{1}{\sqrt{k_1k_2}}
 \tan^{-1}\left[\frac{\sqrt{\frac{k_1}{k_2}}(M-M_0)}{1+\frac{k_1}{k_2}M M_0} \right] .
 \label{k30}
\end{eqnarray}

Since $\frac{k_1}{k_2}M M_0 \gg 1$, alternatively, one can write Eq.~\eqref{k30} as
\begin{eqnarray}\nonumber
 t_{\rm d}\big|_{k_3 \rightarrow 0} & \approx & t_0+\frac{1}{\sqrt{k_1 k_2}}\tan^{-1}\left[\sqrt{\frac{k_2}{k_1}}\left(\frac{1}{M_0}-\frac{1}{M}\right) \right] 
 \\
 &\approx & t_0+\frac{1}{\sqrt{k_1 k_2}}\tan^{-1}\left[\frac{1}{M_0} \sqrt{\frac{k_2}{k_1}} \right] \hspace{.3cm} ({\rm for} ~~ M \gg M_0).
 \label{des1}
\end{eqnarray}

Now considering accretion as the only relevant growth mechanism for EBH and neglecting DM capture, such that
$\dot{m}_{\rm DM} \ll \dot{m}_{\rm acc}$ (equivalently, $\sqrt{k_2/k_1} \ll M_0$), the characteristic timescale $t_{\rm d}$ can be evaluated by setting $k_2 \rightarrow 0$ in Eq.~\eqref{des1}, which yields

\begin{eqnarray}
 t_{\rm d}\big|_{k_2 \ra k_3 \ra 0} \equiv t_{\rm d}^{\rm earlier} = t_0+\frac{1}{k_1}\left(\frac{1}{M_0}-\frac{1}{M}\right) \approx t_0+\frac{1}{k_1M_0}=t_0+ \frac{c_s^3 R^3}{3 G^2 M M_0} \equiv t_0+ t_{\rm acc}^{\rm earlier}
 \label{des2}
\end{eqnarray}
using the Taylor expansion of the square bracket term in Eq.~\eqref{des1}. Here, $t_{\rm acc}^{\rm earlier}= c_s^3 R^3/(3 G^2 M M_0)$~\cite{Chakraborty_low_mass_nakedsingularity2024, GenoliniSerpicoTinyakov2021,mcdermott_constraints_2012,baumgarte_max_surv_2021} and  $t_0$ is the EBH formation timescale (Eq.~\eqref{formation}) as discussed in Sec.~\ref{sec3} and Sec.~\ref{sec4}. Eq.~\eqref{des2} shows that our calculation in Sec.~\ref{sec4} is consistent with the earlier results obtained in~\cite{mcdermott_constraints_2012, Chakraborty_low_mass_nakedsingularity2024, baumgarte_max_surv_2021, GenoliniSerpicoTinyakov2021} for the calculation of the accretion timescale as $t_{\rm acc}^{\rm earlier}$. In our calculation, $t_0$ is also taken into account (with the exact expression of $t_{\rm acc}$) to calculate $t_{\rm d}$ in Eq.~\eqref{td} in Sec.~\ref{sec4}. 
Note that the above approximation always applies in the Bondi regime because the quantum accretion rate becomes relevant only when the EBH mass is small, at which point $\dot{m}_{\rm DM}\sim\dot{m}_{\rm acc}$ and perhaps even $\dot{m}_{\rm Hr}$ are of similar order. Thus, quantum corrections (see Sec.~\ref{sec3.5}) do not alter the above approximation.

\subsection{\label{ss2}Special case: $|\dot{m}_{\rm Hr}| \ll \dot{m}_{\rm acc}$ with $M_0 \ll\sqrt{k_2/k_1}$}
It is well known that the rate of Hawking radiation decreases ($ \sim k_3/M_0^2$) with increasing mass of the BH. Now, if the newly-formed EBH is so heavy that the third term of the right hand side of Eq.~\eqref{tot1} is negligible compared to the first two terms, one should obtain Eq.~\eqref{des1} that reduces to
\begin{eqnarray}
 t_{\rm d} = t_0+\frac{\pi}{2\sqrt{k_1 k_2}}={\rm const.}
 \label{con}
\end{eqnarray}
for $M_0 \ll\sqrt{k_2/k_1}$. Eq.~\eqref{con} is independent of $M_0$, which reveals an interesting situation that could, in fact, arise in reality. 
For example, if an NS captures bosonic DM particles of $m_{\chi} \sim 10^6$ GeV with $\sigma_{\rm n\chi} \sim 10^{-40}$ cm$^2$ in the Galactic bulge, the value of $k_1=2.67 \times 10^{-27}$ kg$^{-1}$.s$^{-1}$ and $k_2=1.26 \times 10^{4}$ kg.s$^{-1}$. An EBH of $M_0=8.54 \times 10^{10}$ kg is formed at $t_0 \approx 113$ yr. As $M_0 \ll \sqrt{k_2/k_1}=2.17 \times 10^{15}$ kg, one directly obtains from Eq.~\eqref{con},
\begin{eqnarray}
 t_{\rm d} = t_0+\frac{\pi}{2\sqrt{k_1 k_2}} \sim 113+8600 \, {\rm yr} \approx 8713 \,{\rm yr}={\rm const.}
\end{eqnarray}
which is also clear from the solid magenta curve in Panel (b) of Fig.~\ref{NSboson}, as well as other curves corresponding to different $\sigma$ with a plateau region that is nearly independent of $m_\chi$.
The above calculation matches neatly with the results obtained by incorporating quantum corrections as well. For example, with quantum corrections, and numerical integration, one obtains $t_{\rm d}=8713$ yr for the above-mentioned parameters. Quantum corrections do not alter the trend significantly but can change the absolute values by a few digits, mostly in the same order of magnitude.
Note that the Hawking radiation rate is negligible compared to the accretion and capture rates in this particular case.

\subsection{\label{ss1}Special case: $\dot{m}_{\rm DM} \ll \dot{m}_{\rm acc}$ and $\dot{m}_{\rm DM} \ll |\dot{m}_{\rm Hr}|$ }

If the DM capture rate ($\dot{m}_{\rm DM}$) is negligibly small (i.e., $k_2 \to 0$) compared to $\dot{m}_{\rm acc}$ and $\dot{m}_{\rm Hr}$, one obtains from Eq.~\eqref{td} or Eq. (\ref{td-mcrit}),

\begin{align}
    t_{\rm d}\big|_{k_2 \ra 0} =t_0+\frac{M_{\rm crit1}}{(k_3+k_1 M_{\rm crit1}^4)}
    \left[\sqrt{k_3/k_1}\,\tan^{-1}\left\{\frac{\sqrt{k_3/k_1}\,M_{\rm crit1}(M-M_0)}{k_3/k_1+M_{\rm crit1}^2MM_0}\right\}-M_{\rm crit1}^2\tanh^{-1}\left\{\frac{M_{\rm crit1}(M-M_0)}{M_{\rm crit1}^2-MM_0}\right\}\right].
    \label{tdk20}
\end{align}

For a consistency check, one may take the limit $k_3 \to 0$ in Eq.~(\ref{tdk20}), which directly reproduces Eq.~(\ref{des2}). Here too, we note that the above result is valid even if quantum corrections are taken into account. Numerically integrating a function that properly accounts for both the Bondi and Unruh regimes still agrees with the timescale obtained from Eq.~\eqref{tdk20}.

\bibliography{MyLibrary}
\bibliographystyle{apsrev4-2}

\end{document}